\documentclass[12pt,preprint]{aastex}

\slugcomment{revised version}

\begin{document}{}

\def\t0{\theta_{\circ}}
\def\muo{\mu_{\circ}}
\def\sd{\partial}
\def\be{\begin{equation}}
\def\en{\end{equation}}
\def\bv{\bf v}
\def\bvo{\bf v_{\circ}}
\def\ro{r_{\circ}}
\def\rhoo{\rho_{\circ}}
\def\etal{et al.\ }
\def\msun{\,M_{\sun}}
\def\rsun{\,R_{\sun}}
\def\lsun{L_{\sun}}
\def\msunyr{M_{\sun} yr^{-1}}
\def\kms{\rm \, km \, s^{-1}}
\def\mdot{\dot{M}}
\def\Md{\dot{M}}
\def\curf{{\cal F}}
\def\ecs{erg cm^{-2} s^{-1}}
\def \haebe{HAeBe}
\def \mum {\,{\rm \mu m}}
\def \simali {{\sim\,}}
\def \K {\,{\rm K}}
\def \Angstrom     {\,{\rm \AA}}

\title{Dust Processing in Disks around T Tauri Stars}

\author{B. Sargent\altaffilmark{1},
W.J. Forrest\altaffilmark{1},
P. D'Alessio\altaffilmark{2},
A. Li\altaffilmark{3},
J. Najita\altaffilmark{4},
D.M. Watson\altaffilmark{1},
N. Calvet\altaffilmark{5},
E. Furlan\altaffilmark{6},
J.D. Green\altaffilmark{1},
K.H. Kim\altaffilmark{1},
G.C. Sloan\altaffilmark{6},
C.H. Chen\altaffilmark{4,7},
L. Hartmann\altaffilmark{5}, 
and J.R. Houck\altaffilmark{6}}


\altaffiltext{1}{Department of Physics and Astronomy, University of 
                 Rochester, Rochester, NY 14627;
                 {\sf bsargent@astro.pas.rochester.edu}}
\altaffiltext{2}{Centro de Radioastronomia y Astrofisica, UNAM, 
                 Apartado Postal 3-72 (Xangari), 
                 58089 Morelia, Michoacan, Mexico}
\altaffiltext{3}{Department of Physics and Astronomy, 
                 University of Missouri, Columbia, MO 65211}
\altaffiltext{4}{National Optical Astronomy Observatory, 950 North 
                 Cherry Avenue, Tucson, AZ 85719}
\altaffiltext{5}{Department of Astronomy, University of Michigan, 
                 500 Church Street, 
                 Ann Arbor, MI 48109}
\altaffiltext{6}{Center for Radiophysics and Space Research, 
                 Cornell University, 
                 Ithaca, NY 14853}
\altaffiltext{7}{Spitzer Fellow}

\begin{abstract}
The 8--14$\mum$ emission spectra of 12 T Tauri stars 
in the Taurus/Auriga dark clouds and in the TW Hydrae association 
obtained with the Infrared Spectrograph 
(IRS\footnote{The IRS is a collaborative venture 
between Cornell University and Ball Aerospace Corporation 
funded by NASA through the Jet Propulsion Laboratory and 
the Ames Research Center.}) on board {\it Spitzer} are analyzed.  
Assuming the 10$\mum$ features originate from silicate grains 
in the optically thin surface layers of T Tauri disks, 
the 8--14$\mum$ dust emissivity for each object is derived
from its Spitzer spectrum. The emissivities are fit with 
the opacities of laboratory analogs of cosmic dust.
The fits include small nonspherical 
grains of amorphous silicates (pyroxene and olivine), 
crystalline silicates (forsterite and pyroxene), 
and quartz, together with large fluffy amorphous silicate grains.
A wide range in the fraction of crystalline silicate
grains as well as large silicate grains 
among these stars are found.  
The dust in the transitional-disk objects CoKu Tau/4, 
GM Aur, and DM Tau has the simplest form of silicates, 
with almost no hint of crystalline components and modest 
amounts of large grains.  This indicates that the dust grains 
in these objects have been modified little from their 
origin in the interstellar medium.  
Other stars show various amounts of 
crystalline silicates, similar to the wide dispersion 
of the degree of crystallinity reported for Herbig Ae/Be stars 
of mass $<$\,2.5$\msun$.  
Late spectral type, low-mass stars can have significant fractions
of crystalline silicate grains.  
Higher quartz mass fractions often accompany 
low amorphous olivine-to-amorphous pyroxene ratios.  
It is also found that lower contrast of 
the 10$\mum$ feature accompanies greater crystallinity.
\end{abstract}


\section{Introduction}

It has long been known that T Tauri stars (TTSs) emit 
infrared (IR) radiation in excess of their stellar 
photosphere \citep[e.g.,][]{me66}.  
\citet{coh73} speculated that silicate dust 
in orbit around these stars was responsible for 
this excess emission.  With observations from 
the {\it Infrared Astronomical Satellite} (IRAS), 
it was shown that the 12--100$\mum$ IR excess emission 
from these young stars could arise from dusty 
accretion disks \citep{ruc85}.  
Many different models for this disk emission have been proposed.  
Both \citet{als87} and \citet{kh95} construct disk models 
including both accretion and reprocessing of stellar radiation.  
In order to explain how disk reprocessing can be responsible 
for the IR excesses of most TTSs, 
\citet[henceforth KH87]{kh87} proposed that disks around TTSs are flared, 
in that the scale height of the disk increases more than linearly with the distance from the central star.  
A flared disk intercepts a larger solid angle of radiation 
emitted from the star than a flat or nonflared disk, 
leading to more reprocessing of starlight.  
KH87 suggest that the surface of such a flared disk 
would become hotter than the midplane due to 
radiative transfer effects.  The disk material is optically thick to 
$\lambda$\,$\sim$\,1$\mum$ stellar radiation, 
so the starlight is absorbed in the highest layers of the disk.  
At $\lambda$\,$\sim$\,10$\mum$, characteristic of the reprocessed 
radiation from the top disk layers at R\,$\sim$\,1\,AU in the disk, 
the disk has less optical depth, and the reprocessed radiation 
diffuses into the interior parts of the disk and heats those regions.  
For the small accretion rates typical of TTS, the disk atmosphere heats 
to a higher temperature than the layers in the disk underneath, 
and the vertical temperature inversion produces silicate features 
in emission \citep{cal92}.

\citet{dor03} summarizes how, based on the IR spectroscopic 
observations of \citet{gls68}, it came to be established 
that the 10$\mum$ emission (or absorption) feature, 
the broad emission (or absorption) feature from 8 to 13$\mum$ 
seen in a number of astronomical objects, 
is due to the Si--O stretching modes in silicate grains.  
Spectrophotometric observations by \citet{fs76} 
and \citet{fhr76} at wavelengths longer than 16$\mum$ 
provided further support for the silicate hypothesis.  
They found an 18.5$\mum$ peak in the Trapezium emission 
and an 18.5$\mum$ maximum in the absorption from the BN-KL source; 
the pairing of the 18.5$\mum$ feature (the broad emission or 
absorption feature from 16 to 23$\mum$) with the 10$\mum$ features 
in the Trapezium and the BN-KL source confirmed the silicate hypothesis.

There have been many studies of the silicate features of TTSs, 
both ground-based \citep{cw85,hon03,ks05} and space-based \citep{nmb00,ks06}.  
However, ground-based spectroscopic observations are limited by 
the Earth's atmosphere at mid-IR wavelengths in both wavelength 
coverage and sensitivity.  Space-based missions, such as the {\it Infrared 
Space Observatory} \citep[ISO;][]{kess96} do not suffer these limitations.  
The {\it Spitzer Space Telescope} \citep{wer04} offers greater sensitivity 
than previous space-based missions.  Here we focus on studying 
the 10$\mum$ silicate features of TTSs with the Infrared Spectrograph 
\citep[IRS;][]{hou04} on board Spitzer.

It is generally believed that the disks and planetary systems of young stellar objects (YSOs) form from 
material from the ISM. {\it Spitzer} IRS spectra of objects 
with Class I spectral energy distributions (SEDs) \citep{wat04}, 
objects believed to be young protostars still surrounded by 
collapsing envelope material from their parent cloud of gas and dust, 
show smooth, featureless 10$\mum$ absorption profiles, 
indicating amorphous silicates.  
\citet{for04} presented the spectrum of CoKu Tau/4, 
a T Tauri star with a 5--30$\mu$m spectrum well modeled by 
\citet{daless05} by an accretion disk nearly devoid of small 
dust grains within $\simali$10\,AU.  
Unlike the complex 10$\mum$ emission features 
of many Herbig Ae/Be stars indicative of thermally processed silicates 
\citep{bouw01,vb05}, the 10$\mum$ emission feature of 
CoKu Tau/4 is smooth, relatively narrow, and featureless, 
as are the silicate absorption profiles of 
the ISM \citep[e.g.,][]{kemp04} and Class I YSOs \citep{wat04}.  
Smooth, narrow, and featureless profiles indicate amorphous silicates.  
Other objects, such as FN Tau \citep{for04}, have significant 
crystallinity of silicate dust in their disks, evident in the structure in their 10$\mum$ emission features, 
and others still, such as GG Tau A, have larger grains 
as shown by the greater width of the 10$\mum$ emission feature.

In the following, we use optical constants and opacities of various materials 
to model the 10$\mum$ features of our objects.  
We model the dust emission of the six objects whose spectra
are presented by \citet{for04}, TW Hya 
and Hen 3-600 A by \citet{uch04}, 
V410 Anon 13 by \citet{fur05a}, 
and GM Aur and DM Tau by \citet{cal05}; 
we also present and model the 5--14$\mum$ spectrum of GG Tau B.
Stellar properties for our TTS sample are given in Table 1.  
In \S2, we describe our data reduction techniques.  
In \S3, we detail how we derive and fit an emissivity 
for each object, and in \S4 we describe the fit to the derived 
emissivity of each object.  
We discuss our fits in \S5 and summarize our findings in \S6.

\section{Data Reduction}

\subsection{Observations}

The present 12 TTS were observed with the IRS on board {\it Spitzer}
over three observing campaigns from 2004 January 4 to 2004 March 5.  
All objects were observed with both orders of the Short-Low (SL) module 
(R $\simali$60--120; second order [SL2], 
$\Delta$$\lambda$\,=\,0.06$\mum$ [5.2--7.5$\mum$]; first order [SL1], 
$\Delta$$\lambda$\,=\,0.12$\mum$ [7.5--14$\mum$]).  
Fainter objects were observed with the Long-Low (LL) module 
(R$\simali$60--120; second order [LL2], 
$\Delta$$\lambda$\,=\,0.17$\mum$ [14--21.3$\mum$]; 
first order [LL1], $\Delta$$\lambda$\,=\,0.34$\mum$ [19.5--38$\mum$]), 
while brighter objects were observed with the Short-High 
(SH; R$\simali$600, 9.9--19.6$\mum$) and Long-High 
(LH; R$\simali$600, 18.7--37.2$\mum$) modules 
(the LH spectra are not used here).

The brightest objects were observed in mapping mode, 
in which for one module one data collection event 
(DCE; sampling of spectral signal from target) was 
executed for each position of a 2$\times$3 
(spatial direction $\times$ spectral direction) raster 
centered on the coordinates of the target. 
For details on how the 2$\times$3 maps were obtained, 
see the description of spectral mapping mode by \citet{wat04}.  
From mapping-mode observations, we derive our spectra from 
the two positions in the 2$\times$3 map for which the flux levels 
in the raw extracted spectra are highest.  
All other objects were observed in staring mode, 
which always immediately followed single high-precision 
Pointing Calibration and Reference Sensor (PCRS) peak-up observations.  
For details on IRS staring mode operation and PCRS observations, 
see \citet{hou04}.  For staring mode, the expected flux density 
of the target determined the number of DCEs executed at one 
pointing of the telescope; for faint objects, multiple DCEs 
were obtained at one pointing of the telescope and averaged together.

From the position of each target's point-spread function (PSF)
in the cross-dispersion direction in the two-dimensional data, 
we conclude that mispointing in the cross-dispersion direction in 
SL in mapping mode and in staring mode is usually less
than 0$\farcs$9 (half a pixel).  By comparing the absolute 
flux levels of the spectra obtained from each of the three positions 
in a 1$\times$3 subsection (the three positions are colinear and 
offset from each other in the dispersion direction of the slit) 
of the 2$\times$3 raster, the pointing of the telescope in 
the dispersion direction in mapping mode could be determined.  
The mapping mode dispersion direction mispointing is usually 
less than half a pixel in SL (0$\farcs$9).  Pointing in 
the dispersion direction cannot be quantified very easily for 
staring mode observations because the telescope is 
not moved in the dispersion direction in this mode.  To account for mispointing in 
the cross-dispersion direction, we use data that 
has been divided by the flatfield from the S11 pipeline.  
Mispointing in the dispersion direction primarily causes small photometric error; this will not affect the emissivities derived here.

\subsection{Extraction of Spectra}

The spectra were reduced using the Spectral Modeling, Analysis, 
and Reduction Tool \citep[SMART;][]{hig04}.  
From basic calibrated data (BCD; flat-fielded, 
stray-light-corrected, dark-current-subtracted) 
S11.0.2 products from the {\it Spitzer} Science Center 
IRS data calibration pipeline, permanently bad (NaN) pixels 
were fixed in our two-dimensional spectral data.  
The corrected pixel value was a linear interpolation 
of the nonbad pixels of the set of four nearest neighboring pixels 
(up, down, left, and right of the pixel in question).  
Unresolved lines of [Ne {\footnotesize II}] and molecular hydrogen, seen 
in the spectra of other objects, provided the wavelength 
calibration; all spectra presented in this paper are
 wavelength-corrected, and these wavelengths are 
estimated to be accurate to $\pm$\,0.02$\mum$.  
All DCEs taken at the same pointing of the telescope 
(same module/order/nod position) were averaged together. 
Because one order records the spectrum of sky 
$\simali$1\arcmin--3\arcmin away from the target
 whose spectrum is being recorded in the module's 
other order, sky subtraction in low resolution spectra 
obtained in staring mode is accomplished by subtracting 
the average spectrum from one spectral order of 
a given module from that in which the spectrum is 
located in the same nod position of the other order 
of the same module.  For FM Tau, the SSC pipeline 
introduced artifacts in the off-order images in SL2.  
In this case, we used the same-order, 
different-nod DCE to subtract the sky.

The low-resolution sky-subtracted spectra were then 
extracted using variable-width column extraction in order 
to account for the linear increase of size of 
the object's PSF with wavelength.  From shortest to 
longest wavelengths of each module, respectively, 
extraction region width varied from 3.2 to 4.9 pixels 
in SL2, 3.6 to 4.4 pixels in SL bonus order 
(a short fragment of first-order light from 
7.5--8.4$\mum$ recorded simultaneously with SL2), 
2.7-5.4 pixels in SL1, 3.2-4.9 pixels in LL2, 
3.8-4.3 pixels in LL bonus order (a short fragment 
of first-order light from 19.4 to 20.9$\mum$ recorded
simultaneously with LL2), and 2.5 to 5.5 pixels in LL1.  
For SH the sky is not subtracted, as the SH slit is only 
5 pixels long, and no separate sky observations were acquired.  
The extraction region at each wavelength for SH was
the entire 5 pixel long slit.  Since the roughly square 
SH pixels are $\simali$2$\farcs$2 wide 
and SL pixels are $\simali$1$\farcs$8 wide, 
the SH extraction region covered more solid angle 
at every wavelength than SL, 
and as spectra from SH typically gave lower flux 
than SL over the wavelength range of overlap of SH 
with SL ($\simali$10--14$\mum$), the sky levels for 
the SH observations are estimated to be much lower 
than the flux density of the point sources.

The spectra are calibrated using a relative spectral 
response function (RSRF), which gives flux density, 
F$_\nu$, at each wavelength based on the signal detected 
at that wavelength.  The RSRFs were derived by dividing 
the template spectrum of a calibrator star by the result 
of extraction of the calibrator's spectrum in SMART for 
each nod of each order of each module.  For both orders 
of SL and for both orders of LL, a spectral template of 
$\alpha$ Lacertae (A1\,V; M. Cohen 2004, private communication) 
of higher spectral resolution than the templates described 
by \citet{coh03} was used, and a spectral template for 
$\xi$ Dra (K2\,III) was used for SH \citep{coh03}.  
As with the science targets, SH observations of 
the calibrator source $\xi$ Dra were not sky-subtracted.  
The science target raw extractions were then multiplied 
by the RSRFs corresponding to the same nod, order, and module.  
Typically, good flux agreement at wavelength regions of 
order overlap within the same module was found.  
As the spectra obtained for a given source at the telescope's 
two nod positions are independent measurements of 
the object's spectrum, close agreement between the two nod 
positions was expected; 
this was the case for all sources except for FN Tau.

\subsection{Remaining technical problems}

FN Tau was observed in mapping mode, and from the extracted flux 
levels of all observations in the 2$\times$3 mapping raster, 
we determined that the central 1$\times$2 pair both suffer 
mispointing of differing amounts.  The more mispointed DCE was 
mispointed in the dispersion direction by 0$\farcs$7, 
while the less mispointed DCE was mispointed in 
this direction by 0$\farcs$3; both DCEs were mispointed 
in the cross-dispersion direction by 
between 0$\farcs$5 and 0$\farcs$7.  
The effect of this differential mispointing shows up 
most prominently in the derived spectrum of the more 
mispointed observation of first order of SL; 
there is a mismatch of flux level of about 10$\%$ 
over the entire order compared to the flux level of 
the spectrum obtained from the less mispointed DCE.  
To correct for this, the first order of the spectrum from 
the more mispointed mapping position was multiplied 
by 1.1 to match the less mispointed position.  
Except for the first order of SL for FN Tau, 
the derived spectra are the mean at each order of 
each module of the spectra from the two independent 
nod or map positions.  For the first order of SL for FN Tau, 
the reported spectrum is that from only 
the less mispointed map position.  
Error bars are derived for each of the spectra, 
and the error bar at a given wavelength is equal 
to half the difference between the flux (at that wavelength)
from the two nod (or map) positions 
used to derive the mean spectrum.  
For SL first order of FN Tau, the spectrum from the less 
mispointed map position and the corrected spectrum 
(previously described) from the more mispointed map 
position were used to derive its error bars.  
Any error bar with relative uncertainty $<$1\% is attributed 
to the low number (2) of measurements at that wavelength, 
and that errorbar is set to 1\% of the flux.

There are some mismatches in flux between SL and SH, 
and between SL and LL in the spectra.  Comparing the SL spectra 
of nonvariable sources to available photometry, 
absolute spectrophotometric accuracy is estimated 
to be better than 10\% in SL.  
Therefore, small mismatches in flux levels between 
SL, SH, and LL are corrected by scaling the entire 
longer wavelength module to match the flux in SL, 
as we trust the photometric levels of SL.  
SH was multiplied by factors between 1.04 and 1.11 
to match SL; LL for CY Tau was multiplied 
by 0.95 to match its SL spectrum.  In order to account for 
off-order leaks in the filters which define the orders of 
each of the modules, the ends of each order of every module 
are truncated to guarantee the spectral purity of our spectra.  
The spectra of all objects in the sample excluding GG Tau B 
have been previously published: CoKu Tau/4, FM Tau, IP Tau,
GG Tau A, FN Tau, and CY Tau in \citet{for04}; 
TW Hya and Hen 3-600 A in \citet{uch04}; 
V410 Anon 13 in \citet{fur05a}; 
and GM Aur and DM Tau in \citet{cal05}.  
For all previously published spectra 
except the two by \citet{cal05}, 
wavelengths from 8 to 14$\mum$ were too long by 0.05$\mum$; 
as described previously, this wavelength problem was corrected 
before further analysis.  The correction has moved 
the 9.4$\mum$ feature in FN Tau noted by \citet{for04} closer
to 9.3$\mum$. In Figure 1, the spectrum obtained of the GG Tau B binary system is shown; 
in \S4 we discuss the origin of the IR excess for this pair.

\section{Analysis}

\subsection{Correction for Extinction}

For all of the objects in the sample except V410 Anon 13, 
no correction for extinction is applied, in order not to 
introduce artifacts of overcorrection for extinction.  
See Table 1 for the assumed visual extinction $A_V$ for 
each of the objects in our sample.  
No extinction correction is applied for any object 
having $A_V$ less than 1.4; this includes all objects 
in the sample except CoKu Tau/4, GG Tau A, and V410 Anon 13.  
As described by \citet{daless05}, optical spectra of 
CoKu Tau/4 indicate time-dependent reddening to the star.  
This suggests that the source of the extinction to CoKu Tau/4 
may be local to the object.  However, the precise time and space 
dependence of extinction to CoKu Tau/4 is unknown.  
For this reason, no extinction correction is applied for CoKu Tau/4.  
\citet{wgrs99} estimate $A_V$ of $\sim$ 3.2 toward GG Tau Ab 
(the GG Tau system is a hierarchical quadruple with 
the northern pair, GG Tau A, being binary and separated 
by 0$\farcs$25 -- this is described in greater detail in \S4), 
while $A_V$\,$\approx$0.72 toward GG Tau Aa.  
Consequently, no extinction correction is applied for GG Tau A, as it is believed the less extinguished GG Tau Aa component 
dominates the IRS spectrum.

For V410 Anon 13, an extinction correction is applied, 
assuming $A_V$\,$\approx$5.8 along with \citet{fur05a}.  
Furlan \etal present a disk model for this object to fit its IRS spectrum assuming an inclination
$i$ = 70\degr.  Because of the large implied disk inclination, 
at least part of the extinction to V410 Anon 13 could be 
due to dust in the flared disk atmosphere at large disk radii 
lying in the sightline from the star and inner disk regions 
to observer.  \citet{fur05a} also found that when $i$ 
in the model is changed from 70\degr to 60\degr, 
the peak of the flux in the 10$\mum$ feature increases by 
$\simali$20$\%$.  Because the emergent disk spectrum in 
the model greatly depends on its inclination $i$, 
the effect of extinction correction for this object is 
discussed in \S4.12 when describing its dust model fit.  
Because dust in the outermost reaches of YSO disks is 
expected to be little altered from its origin in the ISM 
(see the discussion in \S5 below), 
the composition of any dust providing local extinction is 
assumed to be approximately the same as that in the ISM 
between V410 Anon 13 and Earth.

In the ISM, the ratio of the visual extinction to 
the optical depth at the 9.7$\mum$ peak of 
the silicate absorption feature ($A_V$/$\Delta$$\tau_{9.7}$)
varies by as much as a factor of 2 to 3 (see Draine 2003). 
To convert from $A_V$ to the 9.7$\mum$ extinction,
we take $A_V$/$\Delta$$\tau_{9.7}$\,$\approx$18, typical for
the local diffuse ISM (see Draine 2003).  
For simplicity, we assume that the composition of 
the material responsible for the extinction does not 
change over the sightline from the target to Earth.  

\subsection{Derivation of Emissivity}

The spectral excess for each of the objects in the sample 
is interpreted as arising from a disk surrounding one or 
more central star(s) beginning in most cases at a few 
stellar radii away from the central star(s) and extending 
as far away as a few hundred AU.  
We call a disk a ``transitional disk'' if it is 
optically thick to mid-IR wavelengths over some range 
of radii and optically thin elsewhere. 
CoKu Tau/4 \citep{daless05}, DM Tau \citep{cal05}, 
GM Aur \citep{cal05}, TW Hya \citep{cal02}, 
and Hen 3-600 A \citep{uch04} have been shown to 
be transitional disks through spectral modeling.  
The spectrum of CoKu Tau/4 is photospheric at wavelengths 
shortward of 8$\mum$ but has a large IR excess seen in its
IRS spectrum longward of that wavelength; 
correspondingly, it has an optically thick 
disk at radii greater than 10\,AU with less than 
$\simali$0.0007 lunar masses of small silicate dust grains 
inside that radius \citep{daless05}.  DM Tau has an IRS spectrum similar 
to that of CoKu Tau/4 and is modeled similarly by \citet{cal05}, 
but with the radius of transition between the optically thick 
disk and the (very) optically thin inner regions at 3\,AU.  
GM Aur, TW Hya, and Hen 3-600 A, also with large excess 
above photosphere longward of 8$\mum$, are not photospheric 
shortward of 8$\mum$.  This excess indicates an optically thin 
inner disk region.  The IR disk emission for each of 
the transitional disks is isolated by subtracting 
an appropriate stellar photosphere represented by 
a blackbody.  For CoKu Tau/4 and DM Tau, the blackbody 
is fit to the 5--8$\mum$ IRS spectral data, 
while for GM Aur, TW Hya, and Hen 3-600 A the Rayleigh-Jeans 
tail of the blackbody is fit to that of the stellar photosphere 
model by \citet{cal05} and \citet{uch04}, respectively.  
This isolated disk emission for each object should, 
shortward of $\simali$20$\mum$, be due in large part 
to emission from the optically thin regions of the disk.  
For CoKu Tau/4 and DM Tau, this emission is mostly due to 
the optically thin regions of each object's wall.  
For GM Aur, TW Hya, and Hen 3-600 A, this emission is 
mostly due to the optically thin inner disk regions \citep{cal05,uch04}.

We refer to a ``full disk'' if the disk is optically thick 
to mid-IR wavelengths throughout and extends from the dust-sublimation radius from the central star.  
Following the reasoning of \citet{for04}, 
FM Tau, IP Tau, GG Tau A, GG Tau B, FN Tau, 
V410 Anon 13, and CY Tau are identified as having 
full disks based on their 5--8$\mum$ spectra.  
Each has a continuum from 5 to 8$\mum$ characterized 
by a spectral slope shallower than the Rayleigh-Jeans 
tail from a naked stellar photosphere.  
In addition, the 5 to 8$\mum$ flux exceeds 
that from stellar photosphere alone (modeled by 
fitting a stellar blackbody to the near-IR photometry) 
by factors $>$2.  
Following the discussion by \citet{for04}, 
most of the 5--8$\mum$ emission from full disks 
originates from optically thick inner disk regions, 
while most of the emission in the dust features above 
the continuum longward of 8$\mum$ is due to emission 
from dust suspended in the optically thin disk atmosphere. 
Therefore, a power law continuum is fit to the $<$8$\mum$ 
region of each ``full-disk'' spectrum and subtracted from 
the spectrum to isolate the optically thin disk atmospheric emission.

Dust grains suspended in the optically thin atmosphere of 
a flared disk are directly exposed to stellar radiation, 
which heats the grains above the temperature of the disk's photosphere.  The grains then reemit 
the absorbed energy according to their temperature; 
this emission gives rise to the distinctive dust features 
seen in the spectra beyond 8$\mum$.  The emission features 
are much narrower than a Planck function,
which indicates structure in the dust emissivity.  
As explained in \citet{cal92}, a radiatively heated 
disk with a modest accretion rate has a thermal inversion.  
The upper layers of the disk, the optically thin disk atmosphere, 
are hotter than the lower layers, which are optically thick. 
This gives rise to spectral emission features characteristic 
of the dust in the disk atmosphere.

In modeling the SEDs of Classical T Tauri Stars (CTTSs), 
both \citet{cal92} and \citet{daless01} compute the temperature 
of the atmosphere of each annulus of disk material as
a function of vertical optical depth.  In such models, 
it is assumed that all dust grains at a given height in 
an annulus are at the same temperature, independent of 
grain composition and grain size.  It is similarly assumed 
here that all grains in any sufficiently small volume 
in a disk are at the same temperature, 
independent of grain composition and grain size.  
We aim for a simple model in order to determine 
the composition of the part of the disk giving rise 
to the optically thin dust emission.  
\citet{bouw01} and \citet{vb05} also model dust emission 
by assuming a single temperature for all dust components.  
Optically thin emission from dust over the range of radii 
(and therefore temperatures) that contributes most to 
the 8--20$\mum$ range is represented by emission 
from optically thin dust at a single, 
``average'' temperature, $T$.  
It is assumed that the monochromatic flux of 
this optically thin emission over the short range 
from 8 to 14$\mum$ is given by
\begin{equation}
F_{\nu} = {\Omega_{d}}{\tau_{\nu}}B_{\nu}(T) = \epsilon_{\nu}B_{\nu}(T) ~~ ,
\end{equation}
where $F_{\nu}$ is either the photosphere-subtracted 
residuals (for transitional disks) 
or power-law-continuum-subtracted residuals (for full disks); 
$\epsilon_{\nu}$ is referred to as the emissivity; 
$\Omega_{d}$ is the solid angle of the region of optically 
thin emission; and $\tau_{\nu}$ is the frequency-dependent 
optical depth of dust.  For all objects except GG Tau B 
and GM Aur, $T$ is found by assuming a long- to short-wavelength emissivity ratio, $\epsilon_l$/$\epsilon_s$, 
with ``$l$'' meaning long wavelengths ($\simali$20$\mum$) 
and ``$s$'' meaning short wavelengths ($\simali$10$\mum$), 
for dust and solving for $T$ in the equation
\begin{equation}
\frac{{F_{\nu}}({\lambda_{l}})}{{F_{\nu}}({\lambda_{s}})} 
= \frac{{\epsilon}({\lambda_{l}})B_{\nu}({\lambda_{l}},T)}
{{\epsilon}({\lambda_{s}})B_{\nu}({\lambda_{s}},T)} ~~.
\end{equation}

For GG Tau B, where no long-wavelength data exist, 
a dust temperature of 252\,K is assumed, 
the same temperature as for GG Tau A.  
For reasons discussed in \S4.4, the dust temperature is 
set to $T$\,=\,310\,K for GM Aur.  
For all other objects observed with SL and LL excluding DM Tau, 
we take the 20$\mum$-10$\mum$ flux ratios; 
for the objects observed in SL, SH, and LH, 
we decrease the wavelengths in the ratio to 19.3 and 9.65$\mum$, 
as SH does not extend to 20.0$\mum$.  
In a single-temperature dust model, 
the same temperature will be computed regardless 
of the wavelengths used to determine the flux ratio 
due to the properties of the Planck function.  
The wavelengths were set to 9.5 and 19.0$\mum$ for DM Tau
as a test to determine if the derived temperature depended 
much on the exact choice of wavelengths used 
to the determine flux ratio.  
When the wavelengths were changed to 10 and 20$\mum$ for DM Tau, 
the computed temperature changed from 160 to 158\,K; 
however, this did not require any change to the DM Tau dust model.  
A similar test was performed on FN Tau by changing the long and 
short wavelength fluxes used for its dust temperature determination 
from 10 and 20$\mum$ to 9.5 and 19$\mum$.  
This increased dust temperature from 208 to 209\,K; 
as with DM Tau, no change to the FN Tau dust model was required.

Using this temperature, $T$, 
the photosphere- or continuum-subtracted residuals 
were divided by $B_\nu(T)$ to give the emissivity, 
which is proportional to the mass-weighted sum of opacities as follows:
\begin{equation}
\epsilon(\lambda) 
\propto \sum_j m_{j}\kappa_{j}(\lambda) = \sigma(\lambda) ~~,
\end{equation}
where $m_{j}$ is the mass fraction of dust component $j$, 
$\kappa_{j}$($\lambda$) is the wavelength-dependent opacity 
(cm$^2$\,g$^{-1}$) of dust component $j$, 
and $\sigma$($\lambda$) is the wavelength-dependent 
cross-section of the dust mixture model.  
Both $\epsilon$($\lambda$) and $\sigma$($\lambda$) 
are normalized to unity at their peak in the 8--14$\mum$ range.

To determine the uncertainties in the emissivities, 
the corresponding spectral error bars, 
obtained as described in \S2.3 from spectra obtained 
at two nod positions, are divided by $B_\nu(T)$, 
the result of which is then divided by the same 
normalization constant used to derive
the corresponding emissivity.  
The derived emissivities are believed to be valid 
immediately longward of 8$\mum$, where the 10$\mum$ feature 
rises above the extrapolation of the $<$\,8$\mum$ continuum, 
as no drastic change of the slope of the continuum from 
the optically thick components of the disks at wavelengths 
between 8 and 14$\mum$ is expected.  However, assuming one 
dust temperature for a wide range of wavelengths in 
a spectrum of a circumstellar disk is unrealistic.  
In addition, for wavelengths longward of $\simali$14$\mum$, 
the slope of the continuum from optically thick emission 
is not well determined by extrapolation from 
the $<$8$\mum$ continuum.  
The power-law-continuum- or photosphere-subtracted residual flux at $\simali$20$\mum$, attributed as described previously 
to optically thin emission, is therefore uncertain, leading to 
uncertainty in the derived dust temperature.  For this reason, 
we do not attempt to fit the 18$\mum$ and longer wavelength features here. 
The emissivity is fit by finding the optimal set of mass fractions, 
$m_{j}$, such that the normalized model dust cross-section fits 
the normalized emissivity as well as possible.  
The fitting method is iterated until the assumed 
$\epsilon_l$/$\epsilon_s$, used to compute grain temperature 
and therefore derive emissivity, 
equals the $\epsilon_l$/$\epsilon_s$ derived 
from the fit emissivities; $\epsilon_l$/$\epsilon_s$ and other details of derivation 
of emissivities are reported for each of the opacity models 
in Table 2.  Also listed in Table 2 is $\beta_{9.9}$, 
the ratio of the continuum-subtracted residual 
flux at 9.9$\mum$ to the 9.9$\mum$ continuum of the full disks, which gives the contrast of the silicate emission 
feature to the optically thick continuum in the original spectrum.

\subsection{Disk Model for IP Tau}

As a test of these simple dust models, a disk model following 
the methods of \citet{daless98,daless99,daless01} was computed, using stellar parameters 
from Table 1 and the mass accretion rate from \citet{har98}.  
First, opacities similar to those generated from the fit to the emissivity 
of IP Tau (see \S4.7) were used to determine the temperature structure of 
the disk, but it was found that the disk would not heat up 
sufficiently--the spectrum generated from this model 
had lower flux than the IP Tau spectrum at all wavelengths.  
A source of dust opacity with high absorption at visible 
and near-IR wavelengths was needed to absorb the stellar 
radiation and heat up the disk to give more flux at all 
IRS wavelengths (5--36$\mum$).  Facing the same issue 
when attempting to model the interstellar extinction, 
\citet{dl84} added in an artificial source of absorption 
for $\lambda$\,$<$\,8$\mum$ in their ``astronomical silicates'' 
for similar reasons (also see Jones \& Merrill 1976;
Rogers \etal 1983).  
Therefore, the optical properties of graphite 
and ``astronomical silicates'' from \citet{dl84}
were used to determine the radial and vertical 
temperature structure of the disk.  Then, opacities obtained 
from the fit to the IP Tau 10$\mum$ feature were used to 
determine the emergent spectrum.  This disk model uses 
the same dust opacities at all radii and at all heights in the disk.  
As can be seen in Figure 2, the 5--12$\mum$ model spectrum is in 
quite good agreement with the observed spectrum of IP Tau.  
It is concluded that the emissivities derived here are basically correct.  
At wavelengths $>$12 $\mu$m, the model gives more flux than is observed.  
There is too much emission from cool grains 
with respect to that from warm grains.  
This may be an indication that dust in the IP Tau disk 
is more settled (settling meaning that the dust-to-gas ratio 
in the disk atmosphere is lower than that at the disk midplane) 
than assumed in the model.  
According to \citet{daless06} and \citet{fur05b}, 
more settling of dust reduces disk emission 
at wavelengths $>$\,20$\mum$ more than shortward of 20$\mum$.

\subsection{Dust Components}

The material in YSO circumstellar disks originates from the ISM.  
\citet{dl84} fit the interstellar extinction with silicates and graphite; 
here, the spectra are also fit with opacities derived from silicates 
and carbonaceous material.  Silicates, which contain SiO$_x$ groups, 
give rise to 10 and 18$\mum$ features from the bonds between Si and O.  
The 10$\mum$ feature arises from Si--O stretching modes, 
while the 18$\mum$ feature is due to O--Si--O bending modes.  
Crystalline silicate grains have many narrow, strong resonances, 
while amorphous silicate grains, lacking ordered structure, 
give rise to two broad, weaker resonances; 
one near 10$\mum$ and one near 18$\mum$.  
From \citet{dor95}, the optical constants for amorphous pyroxene 
of composition Mg$_{0.8}$Fe$_{0.2}$SiO$_3$ and amorphous olivine 
of composition MgFeSiO$_4$ are used.  
In this paper, these dust components are referred to as 
``amorphous pyroxene'' and ``amorphous olivine,'' respectively; collectively, they are referred to as ``amorphous silicates.''  
For optical constants of the three crystalline axes of forsterite 
of composition Mg$_{1.9}$Fe$_{0.1}$SiO$_4$, data from \citet{fab01} 
is used.  The same bulk density, 3.3\,g\,cm$^{-3}$, is assumed for 
forsterite and all small amorphous olivine and pyroxene grains.  This value is obtained from \citet{poll94} for the high end of 
the range of densities for ``high-T silicate''.  
For large amorphous olivine and pyroxene grains, 
this value is the bulk density of the matrix for large porous amorphous 
olivine and pyroxene grains (see \S3.5 for discussion on how porosity modifies the density of a grain).  For crystalline pyroxene, 
mass absorption coefficients for Mg$_{0.9}$Fe$_{0.1}$SiO$_3$ (En90, having Mg/(Mg+Fe) $\sim$ 0.9)
by \citet{chi02} are used.  Chihara \etal report this crystalline pyroxene to have monoclinic structure.  
To represent the opacity profile of silica (SiO$_2$), the dispersion parameters for the ordinary 
ray and extraordinary ray of $\alpha$ quartz by \citet{wc96} are used.  
The density of $\alpha$ quartz assumed is 2.21\,g\,cm$^{-3}$, 
which is the value for amorphous silica stated by \citet{fab00}.  
This is also close to the values of 2.27 and 2.32\,g\,cm$^{-3}$ 
for tridymite and cristobalite (polymorphs of silica), 
respectively \citep{etch78}. Such a value for the density 
of quartz is chosen because it is deemed likely 
(see discussion in \S4) that forms of silica other than 
$\alpha$ quartz are present in the disks studied here, 
possibly in greater mass fractions than $\alpha$ quartz.  
The nominal value for $\alpha$ quartz is 2.65\,g\,cm$^{-3}$, 
as is given in \citet{etch78}.  Because opacity is inversely 
proportional to density, and because the density of quartz
is potentially underestimated here, the silica profile may 
need to be multiplied by a constant slightly less than 1; 
therefore, our mass fractions for quartz could be underestimated 
by as much as a factor of (2.21/2.65)\,$\simali$0.83.  
For the T Tauri sample in this paper, 
we find that adequate fits to the dust emissivity 
do not require large crystalline silicate grains.

As articulated by \citet{dor95}, glassy amorphous silicates 
produced and analyzed in laboratories on Earth are too ``clean'', 
meaning that they do not give enough opacity in 
the 13--16$\mum$ trough between the 10 and 18$\mum$ silicate 
features to account for the dust emissivities of Mira variable 
stars.  Clean silicates also have too little absorption shortward 
of 8$\mum$; Jones \& Merrill (1976) and 
Rogers \etal (1983) argued that 
circumstellar silicates must be ``dirty'' to account 
sufficiently for the absorption of stellar radiation 
by dust in shells around evolved stars.  
The derived disk emissivities in our 
study of T Tauri stars also require a relatively featureless 
source of continuum opacity longward of 13$\mum$.  
The source of this continuum opacity is not well 
constrained because it has no distinctive spectral features.  
We use amorphous carbon to supply the continuum opacity, 
as carbonaceous material is believed to be a major component 
of the ISM \citep[e.g.,][]{dl84, lg97}. 
\citet{zub96} argue in favor of amorphous carbon as a major component of interstellar and circumstellar dust, 
and \citet{brad03} asserts that a large fraction of the carbon 
found in chondritic interplanetary dust particles (IDPs) is amorphous.  
For amorphous carbon, the optical constants for the ``ACAR'' mixture 
\citep{zub96} are used, and a bulk density for amorphous carbon 
grains of 2.5\,g\,cm$^{-3}$ \citep{lis98} is assumed.

\subsection{Dust Shape and Size} 
In addition to grain composition, grain shape and grain size 
affect the computed dust opacities. 
The presence of interstellar polarization indicates 
that interstellar grains must be both elongated and aligned 
(Draine 2003).
\citet{dyck73} observed in the Becklin-Neugebauer (BN) source 
in Orion a correlation of the linear polarization with
the optical depth of the 10$\mum$ silicate band,
leading them to propose the existence of nonspherical aligned 
silicate grains in front of BN.  For an approximation of 
actual grain shapes, we assume here a CDE2 shape distribution \citep{fab01} 
for small amorphous silicates and amorphous carbon 
dust in our models.  In CDE2, all ellipsoidal shapes 
are included, but the distribution is peaked toward 
near-spherical shapes, and extreme shapes such as needles or 
flat sheets are given zero weight.  
CDE2 requires all grains to be in the Rayleigh limit (2$\pi$a/$\lambda$ $\ll$ 1, where a is the radius or characteristic size of grain).
This condition is generally met by submicron-sized grains
for the IRS wavelengths ($\lambda$\,$>$\,5$\mum$). 
For the shape distributions of quartz and forsterite, 
the continuous distribution of ellipsoids \citep[CDEs;][]{bh83} 
is used, in which all shapes are equally weighted, 
because this shape distribution fits the observed dust features 
in the emissivities better than the CDE2 shape distribution.
We note in the Appendix that 0.1$\mum$ sized porous grains of 
$\alpha$ quartz and forsterite give rise to opacity profiles 
nearly identical to those of solid grains of $\alpha$ quartz 
and forsterite with the CDE distribution, respectively.  
We do not suggest that the small crystalline grains are porous, 
but we note this identity.  For forsterite, it is assumed that 
the three crystalline axes are randomly oriented with respect 
to the three ellipsoidal axes.  As with CDE2, CDE also assumes 
particles much smaller than the wavelengths of interest.  
Because opacities for crystalline pyroxene grains ground 
in an agate mortar \citep{chi02} are used, no definitive 
statements about the grain shape distribution for this 
species of dust can be made.  As described by \citet{sog99}, 
forsterite was ground into powder; when observed with a scanning 
electron microscope, the forsterite particles constituting 
the powder are observed to be nonspherical.  
The only statement that can be made about these particles' 
sizes is that they are submicron \citep{sog99}.  
The same is assumed to be true for the crystalline 
pyroxene grains described by \citet{chi02}.

Grain growth is accounted for by including opacities 
derived from grains of a single size to represent large grains, 
following \citet{bouw01} and \citet{vb05}; 
see those references for descriptions of the sizes and shapes 
of the grains they used.  
Mie theory \citep{bh83} can be used to obtain exact results for the opacity of solid 
spherical grains of arbitrary size; 
however, large grains grown by coagulation in 
protoplanetary disks are expected to have 
an irregular and porous structure 
(see Li \& Lunine 2003 and references therein).
To account for this heterogeneity of large grains, 
the Bruggeman effective medium theory (EMT; Bohren \& Huffman 1983)
is used, following Li \& Lunine (2003), in deriving the effective 
dielectric functions of large fluffy grains.  
\citet{lis98} found that using these effective dielectric 
functions in Mie theory could approximate the absorption 
efficiency of grains with fractal dimension $D$, 
a measure of the porosity (volume fraction of empty space) 
in a given grain, between 2.5 and 3; $D$\,=\,3 for solid grains, while
$D$\,$<$\,3 represents porous grains.  
The effective dielectric functions computed from 
the Bruggeman EMT are used in Mie theory 
to derive the absorption efficiencies of 
spherical porous grains.
\citet{har02} use the EMT with 2.5\,$<$\,$D$\,$<$\,3
to model the fluffy grains in comet Hale-Bopp 
(also see Li \& Greenberg 1998; Lisse et al.\ 1998).  
The optical depth through the center of the grains 
is preserved, in order to match the opacity profile of the solid 2$\mum$ amorphous 
silicate grains used by \citet{bouw01}.  Optical depth ($\tau_{\lambda}$) and bulk silicate opacity ($\kappa_{\lambda}$) are kept the same for both solid and porous grains, assuming the silicate matrix of a porous grain has the same opacity as that of an equivalent-mass solid grain, and the diameter for the porous grain is obtained by $\tau_{\lambda}$\,=\,$\rho\,\kappa_{\lambda}\,d$, 
where $\rho$ is the mass density 
of the grain [equal to the bulk silicate density (that of a solid silicate grain) 
times $(1-f)$, where $f$ is the volume fraction of vacuum].  
For $f$\,=\,0.6, a porous grain of radius 5$\mum$ has the same 
optical depth as that through a solid 2$\mum$ radius sphere.  
The fractal dimension was chosen to be 2.766,
a number which is within the acceptable range for $D$.  
The 8--14$\mum$ opacity profile of these porous
grains with radii $a$\,=\,5$\mum$, $D$\,=\,2.766 and $f$\,=\,0.6
(i.e. 60$\%$ of their volume is vacuum)
is very similar to that of solid spherical 
($f$\,=\,0, $D$\,=\,3) amorphous silicate grains 
of radii $\approx$\,2$\mum$.
In this paper, by ``large grains'' we mean
porous amorphous olivine or pyroxene grains 
of radii $a$\,=\,5$\mum$,
volume fraction of vacuum $f$\,=\,0.6,
and dimension  $D$\,=\,2.766. 

\subsection{Degeneracy among Dust Component Opacities}

Our fits to the disk emissivities are not necessarily unique. 
The degeneracy in the mass fraction of amorphous silicate 
components is larger than for crystalline silicate components.
For instance, much of the broad, smooth opacity profile of 
either of the large silicate grain components overlaps 
with both that of the other large silicate grain component 
and both of the small amorphous silicate opacity profiles.  
There is less uncertainty associated with the two small 
amorphous silicate profiles, as they are narrower and stronger 
than the profiles of the large amorphous silicates.  
There is even less uncertainty regarding the crystalline 
silicate components, as their profiles are even narrower 
and often stronger than any of the amorphous silicate profiles.  
Some crystalline silicate identifications may be confused 
by overlapping bands: the opacity profile of crystalline 
pyroxene has a peak at $\simali$9.3$\mum$ that is close 
in wavelength to the 9.2$\mum$ feature of quartz; 
the prominent $\simali$11.3$\mum$ peak of forsterite 
is very close to the weak 11.2$\mum$ feature of
crystalline pyroxene.  In addition, there are polymorphs 
and forms of silica other than $\alpha$ quartz (see \S3.4).  
\citet{speck97} gives 7.5--13.5$\mum$ extinction profiles for silica; 
they all peak at around 9.1--9.2$\mum$.  
This implies a very large degeneracy between 
the different polymorphs and forms of silica.  
In addition, there is degeneracy between different forms 
and polymorphs of crystalline pyroxene.  
The extinction efficiency of clinopyroxene from 
\citet{koi93} shows a 9.3$\mum$ peak higher than 
the other two major peaks in the 10$\mum$ region.  
If one takes the \citet{chi02} profile of crystalline pyroxene 
(En90, which is monoclinic) and adds silica, 
boosting the 9.3$\mum$ feature, and also forsterite, 
boosting the crystalline pyroxene features 
between 11 and 12$\mum$, 
one obtains a profile with feature strengths 
similar to that reported for clinopyroxene in \citet{koi93}.  Therefore, there is 
degeneracy between clinopyroxene and a mixture 
of silica, forsterite, and the crystalline pyroxene used here (En90).  
The featureless opacity continuum from amorphous carbon 
may be reproduced by silicate grains larger than 
$\simali$20$\mum$ in radius, which give essentially 
featureless continuum for 8--20$\mum$ wavelengths.

\section{Results}

The 10$\mum$ emissivity profiles of the ISM, $\mu$ Cep, 
and three of our transitional disks are compared in Figure 3, 
as described in \S4.1 and Table 3.  In Figures 4, 5, and 6 
the derived emissivities and model fits are shown for \
the 12 T Tauri stars.  The parameters of the model fits 
are detailed in Table 4.

\subsection{Comparison between ISM, $\mu$ Cep, and Transitional Disks}

Figure 3 compares the smoothest, simplest 10$\mum$ silicate 
emission features in the TTS sample to the derived optical depth 
profile of the interstellar absorption towards GCS\,3 in 
the Galactic center from \citet[Fig. 3b]{kemp04} and the 10$\mum$ feature of $\mu$ Cep \citep{sloan03},
a mass-losing M2 Ia supergiant \citep{fmh79}.  
According to \citet{lev05}, the visual extinction $A_V$ for $\mu$ Cep 
is $\simali$2, so no correction for extinction is applied.  
For $\mu$ Cep, the emissivity was derived by first subtracting 
a model photosphere of $T_{eff}$\,=\,3500\,K, log(g)\,=\,0, 
and solar metallicity by \citet{bhaus05}, binned to the {\it ISO} Short Wavelength Spectrometer (SWS) 
spectral resolution.  Following the procedure outlined in \S3.2, 
a silicate temperature of 423\,K was derived, based on the residuals 
at 10 and 18.2$\mum$.  The GCS\,3 profile from \citet{kemp04} 
is displayed unaltered.  In order to 
compare the TTS and $\mu$ Cep profiles with the GCS 3 profile, 
a first-order baseline was fit to each of the emissivities 
at $\simali$8 and $\simali$13$\mum$ and subtracted from 
the corresponding emissivity; each residual is scaled to 
the peak of the GCS 3 profile.  The parameters used for 
this process are listed in Table\,3.  
All five silicate profiles in Figure\,3 are generally smooth.  
The profiles of TTS peak at wavelengths closer to 
the peak of the ISM profile than to that of $\mu$ Cep, 
which peaks at a somewhat longer wavelength.  
The close similarity between the GCS\,3 ISM profile 
and the TTS profiles strongly supports the assertion that the dust 
in YSO disks comes from the ISM.

\subsection{CoKu Tau/4}

The smoothest, least complex T Tauri 10$\mum$ feature 
is that of CoKu Tau/4.  \citet{daless05} model this object 
as a standard flared disk of modest inclination with nearly 
all small grains in the inner 10\,AU cleared: literally, a disk in transition.  
This modeling indicates less than 0.0007 lunar masses 
of grains of ISM grain size ($<$\,0.25$\mum$) remain 
in the inner disk.  This configuration is remarkable 
considering the estimated age of the CoKu Tau/4 system, 
which is only 1--3\,Myr.  The 10$\mum$ silicate feature 
in this model is dominated by emission from the cylindrical 
disk ``wall'' at the 10\,AU truncation radius.  
The outer layers of this wall are roughly isothermal, 
as each point on the wall is $\simali$10 AU from the star; 
as such, the emissivity for CoKu Tau/4, derived by assuming 
all grains are at a single temperature, is a reasonable 
first approximation.  The derived dust temperature for 
CoKu Tau/4 is $\simali$121\,K, which is well within 
the range of temperatures in Figure 4 of \citet{daless05} 
computed using amorphous pyroxene and amorphous olivine grains 
in the optically thin region of the CoKu Tau/4 wall.  
This temperature alone indicates the bulk of the material 
is located $\simali$10\,AU from the star as noted by \citet{for04}.

The 10$\mum$ silicate profile of CoKu Tau/4 is 
quite similar to that of the ISM (see Figure 3).  
A satisfactory fit (Figure 4) is achieved with 
nonspherical small amorphous pyroxene and olivine 
grains with a mass ratio of 3.73:1.  
\citet{kemp04} fit the 10$\mum$ interstellar silicate 
absorption profile toward the Galactic center source GCS\,3
with small spherical amorphous pyroxene and olivine grains 
with a mass ratio of 0.18:1.  
Most of this difference in mass ratio comes 
from the differing adopted shape assumptions.  
The CDE2 grain shape distribution assumed here 
shifts the 9.8$\mum$ peak from spherical amorphous 
olivine grains to 9.95$\mum$ and the 9.3$\mum$ peak 
from spherical amorphous pyroxene grains to 9.4$\mum$.  
In addition, the CoKu Tau/4 profile peaks at a slightly 
shorter wavelength (9.55$\mum$) than 
the GCS\,3 ISM profile (9.6$\mum$).  
This could indicate a slight compositional difference, 
although note the large error bars for $\lambda$\,$<$\,9.55$\mum$ 
in CoKu Tau/4.  The lack of narrow features on top of 
the amorphous feature indicates negligible amounts of 
crystalline grains.  At 1--3 Myr, the dust from 10\,AU 
outward in the CoKu Tau/4 disk shows little evidence for processing.

Although small nonspherical grains of amorphous olivine 
and pyroxene account successfully for the peaks of 
the 10 and 18$\mum$ features, the model with small 
silicate grains does not account for a very small 
excess of emissivity on the long-wavelength side of 
the CoKu Tau/4 10$\mum$ feature.  Optically thick emission 
from the dust would widen the 10$\mum$ feature, 
but it would do so to both the long- and short-wavelength 
sides of the feature.  Larger amorphous silicate grains, 
however, can account for this 11.5--12.5$\mum$ excess.  
Larger grains (see \S3) give greater opacity longward 
of the 10$\mum$ silicate peak, but not shortward.  
Consequently, porous silicate grains of 5$\mum$ radius 
are included 4.3$\%$ by mass in the CoKu Tau/4 
dust model.

While models consisting of only silicates can account for 
the positions and shapes of the 10 and 18$\mum$ features, 
the 12--15$\mum$ opacity continuum from laboratory silicates 
is lower than that in the derived CoKu Tau/4 emissivity.  
Amorphous carbon is used to model this excess continuum opacity (\S3).

\subsection{DM Tau}

DM Tau is also a disk in transition. At disk radii 
interior to 3\,AU, less than 0.0007 lunar masses 
of submicron-sized grains remain (similar to CoKu Tau/4), 
according to \citet{cal05}.  The derived dust temperature 
for DM Tau, 160\,K, is only slightly higher than 
that for CoKu Tau/4 (121\,K), which reinforces 
the conclusion that DM Tau has no inner dust disk (see \S4.2).  
As with CoKu Tau/4, the disk at radii larger than 3\,AU 
is optically thick, and the spectrum from 5 to 8$\mum$ 
is photospheric.  The 10$\mum$ feature is of 
somewhat low contrast to the underlying photosphere, 
but our derived emissivity is well fit with 
small grains of amorphous olivine and pyroxene, 
large 5$\mum$ porous amorphous olivine grains, 
and amorphous carbon.  No crystalline grains are 
indicated by the DM Tau profile (Figure 4).

\subsection{GM Aur}

\citet{cal05} assert that GM Aur is also a transition disk, 
albeit with a slightly more complex radial distribution of 
dust than for CoKu Tau/4 or DM Tau.  Their model includes 
an optically thin inner disk from the dust sublimation radius 
out to 5\,AU containing $\simali$0.02 lunar masses of small dust grains, 
negligible small dust grains between 5 and 24\,AU, 
and a full outer disk beyond 24\,AU.  
The usual method of computing dust temperature 
self-consistently based on the short- to long-wavelength flux ratio 
derived from the photosphere-subtracted residuals results 
in a very poor fit to the derived GM Aur emissivity. 
That method gives $T\simali$210\,K, with 50\% by mass 
small amorphous pyroxene, 33\% small amorphous olivine, 
and 17\% amorphous carbon, resulting in a $\chi^{2}$ 
per degree of freedom (dof) of $\simali$100, 
severely underestimating the emissivity 
for $\lambda$\,$<$\,9$\mum$ and slightly 
overestimating the emissivity longward of 13$\mum$.  
Therefore, a self-consistent dust temperature of 
$T$\,=\,386\,K was determined for GM Aur 
using the fluxes at 9.4 and 18.8$\mum$ of 
the optically thin inner disk component 
from the GM Aur model of \citet{cal05}.  
This gave a somewhat better fit to the computed emissivity.  
The model required $\simali$53\% by mass small amorphous 
olivine and $\sim$47 $\%$ amorphous carbon and 
gave a $\chi^{2}$/dof of 44.4, 
with the dust model overestimating the emissivity 
shortward of the 10$\mum$ feature peak and 
underestimating the emissivity longward of 12.5$\mum$.

The low dust temperature of 210\,K resulted from including emission 
from the very cold wall at 24\,AU in the model of \citet{cal05}.  
The second method excludes the wall emission and gave a dust temperature 
of 386\,K.  The dust model assuming dust at 210\,K underestimates 
the 8$\mum$ emissivity while the model assuming 386\,K dust 
overestimates the 8$\mum$ emissivity (and the opposite effects 
happen at $\simali$14$\mum$).  Therefore, the dust temperature 
is set between 210 and 386\,K to achieve the optimal fit to 
the derived emissivity.  By setting $T$\,=\,310\,K, 
a $\chi^{2}$/dof of $\simali$9.8 
was achieved using the dust mass fractions listed in Table 4, 
adequately fitting the computed 8--14$\mum$ emissivity.  
Amorphous olivine is included to match the $\sim$ 9.8$\mum$ 
central wavelength of the GM Aur 10$\mum$ feature.  
Amorphous pyroxene and amorphous carbon are also included; 
however, no large amorphous silicate grains are indicated, 
as any amount of such grains would make the model 10$\mum$ 
feature wider than that observed.  
Likewise, the spectrum indicates negligible amounts 
of crystalline silicates and quartz.

\subsection{TW Hya}

TW Hya is another suspected transitional disk \citep{cal02,uch04}.  
\citet{cal02} propose that it is a disk partially cleared out 
to $\simali$4\,AU; gas and $\simali$0.5 lunar masses 
of grains of radii  $\simali$1$\mum$ populate the inner 
optically thin disk.  \citet{uch04} found a transition 
radius of 3.3\,AU instead of 4\,AU based on 
the IRS spectrum of TW Hya.  The outer disk is 
assumed to be optically thick.  
The derived dust temperature for TW Hya, 193\,K, 
is low (as with CoKu Tau/4) compared to the dust
temperatures of other stars in our sample, 
which further supports TW Hya relatively lacking 
inner disk material.  The 10$\mum$ feature of TW Hya is smooth, 
like those of CoKu Tau/4 and DM Tau; little crystalline material is indicated.  
The derived emissivity peaks at 9.55$\mum$, 
indicating a high-mass fraction of amorphous pyroxene.  
A fairly substantial large grain content is indicated, 
$\simali$24.5\% by mass.  Approximately 1\% by mass 
of forsterite is indicated by the slight knee at 11.3$\mum$.
TW Hya shows two small peaks at 12.4 and 12.8$\mum$.  
According to \citet{speck97}, various forms of amorphous 
and crystalline silica have features peaking 
between 12 and 13 $\mum$.  Quartz is the only one 
with two peaks at roughly 12.4 and 12.8$\mum$.  
This indicates $\simali$1\% quartz by mass.  
At wavelengths shorter than 8.7$\mum$, however, 
the emissivity rises above the model.  
The data in \citet{speck97} shows that 
other forms of silica may better fit this 
short-wavelength ``shoulder.''  
We intend to investigate this in a future study 
using a larger sample of TTS spectra.

With an age of $\simali$10\,Myr \citep{webb99,wein00}, 
TW Hya has very little crystalline material, 
which is somewhat unexpected.  A number of the disks 
in Taurus from our sample, which like CoKu Tau/4 are 
believed to be 1--3\,Myr old, have significantly greater 
crystalline mass fractions than TW Hya.  
Hen 3-600 A, another system in the TW Hydrae association, 
has much greater mass fractions of crystalline grains, 
making it an interesting counterpoint.  
This issue is discussed further in \S5.

\subsection{FM Tau}

FM Tau has a slightly more complex 10$\mum$ feature than the Taurus transitional disks.  
The 5--8$\mum$ continuum for this object is shallower than, 
and well in excess of, the stellar photosphere, 
implying that the optically thick disk around 
this object extends inward to the dust sublimation 
radius \citep{for04}.  The 10$\mum$ feature of FM Tau 
peaking around 9.6--9.7$\mum$ is narrow. It is fit with 
negligible amounts of large grains. 
Forrest et al.\ (2004) note that the spectrum 
of this object has a ``knee'' at 11.3$\mum$, 
indicating the presence of forsterite.  
The mass fraction of forsterite used in 
the dust opacity model for FM Tau is only $\simali$0.3\%. 
The model spectrum is higher than the emissivity 
at $<$8.3$\mum$.  This is attributed to uncertainty 
in the subtraction of the power law 
(which was fit to the optically thick emission 
immediately shortward of 8$\mum$), 
as the derived emissivity immediately 
longward of 8$\mum$ is more sensitive to 
the absolute level of the power law 
than the emissivity at longer wavelengths.

\subsection{IP Tau}

IP Tau is similar to FM Tau.  \citet{for04} also 
ascribe a full disk extending to the dust sublimation 
radius for this object.  The 10$\mum$ feature peaks 
at $\simali$9.6--9.7$\mum$ and is only slightly wider 
than that of FM Tau, indicating large amorphous olivine grains 
of $\simali$7\% by mass.  The 11.3$\mum$ knee 
of IP Tau is more prominent than that of FM Tau, 
indicating a higher crystalline mass 
fraction of forsterite (2\%) than for FM Tau.

The opacities used to fit the emissivity of IP Tau are very similar to those used in the IP Tau disk model described in \S3.3; there are exceptions.  Instead of using CDE2 as the shape distribution for small amorphous olivine and amorphous pyroxene, the CDE shape distribution was used.  For the same reasons that the assumed shape distribution affected the ratio of amorphous pyroxene to amorphous olivine for GCS 3 and CoKu Tau/4 (see discussion in \S4.2), the use of CDE instead of CDE2 for the small amorphous silicates slightly increased the amorphous pyroxene to amorphous olivine ratio.  Also, 2 $\mu$m radius solid grains of amorphous olivine were used instead of 5 $\mu$m radius porous amorphous olivine grains; however, as noted before, the opacity profile of 5 $\mu$m porous amorphous silicate grains is nearly identical to that of 2 $\mum$ solid amorphous silicate grains.  In addition, graphite was used in place of amorphous carbon to give opacity continuum.

\subsection{GG Tau A}

GG Tau A has the next most complex 10$\mum$ feature.  
GG Tau A is binary, and \citet{wgrs99} estimate masses 
from \citet{bcah98} stellar evolutionary models of 
0.78$\pm$0.10$\msun$ for Aa and 0.68$\pm$0.03 $\msun$ for Ab.  
The separation of the two is 0$\farcs$25 ($\sim$ 35\,AU).  
Our spectrum is the sum of GG Tau Aa and GG Tau Ab. 
Similar to FM Tau and IP Tau, the 5--8$\mum$ continuum 
from GG Tau A is well in excess of the stellar photospheres, 
which led \citet{for04} to conclude that there is 
a full inner disk in GG Tau A.  The circumbinary dust 
disk outside of 35\,AU from the center of mass of 
the GG Tau A pair would be too cold to give rise 
to the 10$\mum$ emission seen from GG Tau A. 
Grains of composition of ``astronomical silicates'' 
from \citet{dl84} at 252\,K in radiative equilibrium 
with a blackbody of radius 2$\rsun$ 
and $T_{\rm eff}$\,=\,4000\,K (representing 
the radiation from one star of the GG Tau A pair) 
would be located $\simali$1\,AU from the star.  
The dust temperature of 252\,K (Table 2) implies that silicate 
grains are located $\simali$1\,AU from either Aa or Ab (or both).  
This is consistent with the finding by \citet{naj03} 
of CO emission originating less than 2\,AU from 
one (or both) of the two stars of the GG Tau A system.  
Following \citet{sp95}, regarding disk partitioning 
in multiple systems, it is assumed that the IR excess 
must originate from a circumstellar disk around Aa or Ab, or both.  
White et al.\ (1999) report H$\alpha$ emission of 
57$\Angstrom$ equivalent width (EW) from Aa but only 
16$\Angstrom$ EW from Ab, suggesting a more substantial 
inner disk around the Aa component.  
However, this disk must be truncated outward 
of a few AU, otherwise the interactions with 
the Ab component would disrupt a larger disk.  
The emissivity peaks at 9.5$\mum$, indicating 
large amounts of amorphous pyroxene.  
Mass fractions of large silicate grains 
similar to TW Hya are indicated,
definitely higher than those for FM Tau and IP Tau.  
A prominent 11.3$\mum$ feature indicates a forsterite mass fraction similar to that for IP Tau.  Quartz is suggested by possible features at 12.4 and 12.8$\mum$.

\subsection{GG Tau B}

The IRS mapping-mode observation of GG Tau A 
also included signal from GG Tau B in the SL slit.  
GG Tau B is about 10$\arcsec$ due south of GG Tau A, 
and it is also a binary \citep{wgrs99}. 
The position angle of the SL slit for each position 
of the GG Tau A map was $\simali$344$^{\circ}$, 
close to the position angle of the separation of GG Tau A and GG Tau B, 
so GG Tau A and B were close to maximally separated
in the Short-Low slits of two of the map positions.  
The spectrum of GG Tau B was extracted from the two 
map positions best centered on its position.  
The Short-Low slit is 2 pixels (3$\farcs$6) wide, 
and the extraction aperture used varies from 
as little as 3.3 pixels (6\arcsec) 
at the shortest wavelength of first order to 
as much as 5 pixels (9\arcsec) at the longest 
wavelength.  
This extraction region was centered on GG Tau B; 
at the longest wavelengths, the edge of the extraction 
region in SL closest to GG Tau A is $\simali$2--3 pixels 
away from the center of the PSF of GG Tau A, 
which is believed to be sufficiently far away 
to minimize contribution of signal from GG Tau A.  
White et al.\ (1999) report the two components 
of the GG Tau B binary, Ba and Bb, to be separated 
by 1$\farcs$48 (207\,AU), 
which is smaller than the Short-Low beam.  
The spectrum shown for GG Tau B therefore 
includes signal from both Ba and Bb, 
as well as from circumstellar disk(s) 
around either or both stars.  
White et al.\ (1999) assign masses 
of 0.12 $\pm$ 0.02$\msun$ for Ba 
and 0.044$\pm$0.006$\msun$ for Bb, 
and report H$\alpha$ EW of $\simali$20$\Angstrom$ for Ba 
and 20--43$\Angstrom$ for Bb.  This puts both Ba and Bb near the hydrogen-burning mass limit.  The 10$\mum$ feature from the GG Tau B pair indicates a higher mass fraction of crystalline 
grains than any of 
CoKu Tau/4, TW Hya, FM Tau, IP Tau, or GG Tau A.  
Because the 10$\mum$ emission from GG Tau B is seen 
and also because the separation between Ba and Bb is 
$\simali$200\,AU, it is believed that a circumstellar 
disk(s) exist(s) around either Ba or Bb, or both.  
It is unclear to which component, Ba or Bb (or both), 
the GG Tau B disk emission belongs.  
Currently, we have no long-wavelength data for GG Tau B, 
so instead of deriving a silicate dust temperature 
for GG Tau B, the same dust temperature for GG Tau A, 
252\,K, is used.  The peak of the 10$\mum$ feature 
is much flatter than that of any of the previous sources, 
with sharp inflections at 9.4 and 11.2$\mum$ defining the plateau.  
Due to the low signal ($\simali$80\,mJy at 10$\mum$), 
the S/N is rather low, so only a rough fit to 
the 10$\mum$ feature was attempted.  
The narrow peak in emissivity at $\sim$9.4$\mum$ 
suggests crystalline pyroxene.  
The plateau shape of the rest of the 10$\mum$ 
feature is due to larger grains and an admixture 
of small crystalline grains of 
other composition--forsterite, in this model.

The emissivity and dust model deviate longward of 12.5$\mum$.  Noting that the flux at SL wavelengths for GG Tau A is roughly 10 times that for GG Tau B, and noting that the extraction aperture is largest for the longest wavelengths of the first order of SL, we attribute the $>$12.5 $\mu$m rise of emissivity above model to contamination of signal from GG Tau A.

\subsection{Hen 3-600 A}
Hen 3-600 is another multiple system with mid-IR emission.
\citet{jay99} resolve the Hen 3-600 system into A and B 
components and measure a separation of components of 
1$\farcs$4 (70\,AU assuming the Hen 3-600 pair 
belongs to the TW Hydrae association).  
In addition, Hen 3-600 A itself is 
a spectroscopic binary \citep{tor03}.  
None of the components of the Hen 3-600 system are 
resolved in any of the instrument slits \citep{uch04}, 
but Jayawardhana et al.\ (1999)  determined that 
the circumstellar disk is associated with the A pair. 
Due to the unknown separation of components of 
the spectroscopic binary, the disk(s) in this system 
cannot be assigned to either of the components.  It is assumed that the A pair are separated by less than 1 AU and that the IRS spectrum for Hen 3-600 A arises from a circumbinary disk around the A pair.  The combined luminosity of the two components 
from \citet{webb99} are listed in Table 1.  
\citet{hon03} and \citet{uch04} report that 
the 10$\mum$ feature for this object indicates 
amounts of crystalline silicates comparable 
to those of amorphous silicates.

The derived dust temperature for 
Hen 3-600 A is $\simali$229\,K, 
higher than for other transitional disks.  
This means that this object has more grains in 
its inner disk than other transitional disks, 
consistent with a transition radius of 1.3\,AU 
(\citet{uch04}), the closest (to the star)
among all transitional disks.  
The 10$\mum$ feature (Figure 6) 
shows a 9--11$\mum$ plateau, 
along with a strong, narrow feature at 9.2$\mum$.  
The 9.2$\mum$ feature indicates quartz 
in the CDE distribution \citep{wc96,speck97}.  
Silica was previously identified by \citet{uch04} 
and \citet{hon03}.  Alpha quartz produces smaller peaks 
at 12.4 and 12.8$\mum$, which may correspond to 
a small feature centered at 12.6$\mum$ in the emissivity.  
Alternatively, other polymorphs and forms of silica give 
single-peaked features at $\simali$12.6$\mum$ \citep{speck97}, 
so it is quite possible that other forms or polymorphs of 
silica may be present in Hen 3-600 A.  
In addition to quartz, a larger mass fraction 
of the material responsible for the continuum opacity 
than for any other objects in this sample is needed to 
fit the 12--14$\mum$ continuum.  
Small amorphous carbon grains are assumed, 
but it could just as well come from very large silicate grains. 
The overall flatness of the 10$\mum$ plateau indicates 
a small amount of crystalline pyroxene.  
Forsterite is indicated by the sharp 11.3$\mum$ edge 
of the plateau, and small and large grains 
of amorphous pyroxene are added to smooth 
the crystalline silicate features between 9.2 and 11.3$\mum$.  The emissivity rises above the dust model from 13 to 14$\mum$, although much less drastically than for GG Tau B.  Such a feature in the emissivity could be due to a dust component other than what we use for Hen 3-600 A.

\subsection{FN Tau}

The derived emissivity of FN Tau also indicates 
a large crystalline silicate mass fraction. 
The plateau top to the 10$\mum$ feature is not flat, 
but it slopes downward to longer wavelengths,
moreso than for Hen 3-600 A.  The derived emissivity of 
FN Tau peaks at 9.3$\mum$, a bit longer than that of Hen 3-600 A.  
It also has minor peaks at 9.9, 10.6, 11.2, and 11.5$\mum$.  
All of these peak positions are highly indicative of 
crystalline pyroxene.  Amorphous pyroxene smooths 
the narrow subfeatures at the top of the 10$\mum$ 
sloping plateau, while simultaneously accentuating 
the 9.3$\mum$ feature.  Approximately 1\% by mass of 
quartz is indicated by the model.  
Like TW Hya, there is a ``shoulder'' at wavelengths 
shorter than 8.7$\mum$ that may indicate a silica type 
other than $\alpha$ quartz.  Like GG Tau B and Hen 3-600 A, 
the dust model underestimates the emissivity at wavelengths 
near 14$\mum$ (see the discussion of GG Tau B).

\subsection{V410 Anon 13}

The V410 Anon 13 emissivity is similar to 
those of FN Tau and Hen 3-600A.  
V410 Anon 13 is a very low-mass star 
\citep[$\simali$0.1$\msun$;][]{fur05a}
of spectral type M\,5.75; quite remarkably, 
it is surrounded by a full, flared accretion disk.  
Like FN Tau, the derived emissivity peaks at 9.3$\mum$, 
indicating crystalline pyroxene as opposed to quartz.  
No clear identification can be made for any features 
(or lack thereof) between 12 and 13$\mum$ due to 
relatively high spectral noise, so a small amount 
of quartz is not ruled out.  
Like Hen 3-600 A, the rest of the 10$\mum$ feature 
is a fairly featureless flat plateau with a sharp 
knee at 11.2$\mum$.  The sharp 11.2$\mum$ knee 
indicates forsterite.  Small grains of amorphous 
pyroxene emphasize the 9.3$\mum$ feature and dampen 
the crystalline features of forsterite and 
crystalline pyroxene along the top of the 10$\mum$ plateau.  
As with FN Tau, the 8.7$\mum$ shoulder and 
the emissivity between 13 and 14$\mum$ are both
underestimated by the dust model.

Because the extinction correction was only applied 
for V410 Anon 13, the effects of changing the extinction 
correction on the dust model for this object were explored.  
As discussed previously, it is believed that any local 
extinction would be very similar to the interstellar extinction, 
so only the amount of extinction was varied.  
When the optical depth of intervening material 
was doubled (from $\tau_{9.7}$\,=\,0.32 
to $\tau_{9.7}$\,=\,0.64), the derived dust temperature 
increased from 256 to 289\,K.  
The inferred mass fraction of amorphous pyroxene 
increased from 48\% to 61\%, quartz decreased from 2\% to 1\%, 
crystalline pyroxene decreased from 11\% to 10\%, 
forsterite decreased from 5\% to 2\%, 
and amorphous carbon decreased from 34\% to 26\%.  
The $\chi^{2}$/dof increased 
slightly from 1.9 to 2.1, 
but the fit still looked reasonable.  
The increase of derived dust temperature 
results from the opacity of interstellar silicates 
being higher in the 10$\mum$ feature than in the 20$\mum$ feature.  
The modest increase in the mass of amorphous pyroxene 
is attributed to the variation of the amount of 
extinction correction over the 10$\mum$ feature.  
The extinction correction is greatest near $\simali$9.7$\mum$, 
vertically ``stretching'' the 10$\mum$ feature 
so that more amorphous pyroxene, peaking at 9.4$\mum$, is needed.  
The extinction correction also has the effect of ``rounding off'' 
any narrow peaks due to crystalline silicates, 
such as the 11.3$\mum$ peak best fit by forsterite, 
thus decreasing the amount of crystalline silicates.  
In addition, this stretching raises the contrast 
of the 10$\mum$ feature above the $>$13$\mum$ continuum, 
effectively decreasing the amount of continuum opacity 
required from amorphous carbon.

\subsection{CY Tau}

The most troublesome emissivity to characterize 
was that of CY Tau.  While the 5--8$\mum$ excess 
of this object is similar to those of the other full 
disks in our sample, the contrast of the 10$\mum$ excess 
to the underlying continuum is the lowest in the entire sample 
(very low $\beta_{9.9}$; see Table 2).  
For this reason, the derived emissivity is poorly 
determined over the entire 10$\mum$ feature.  
What can be discerned is a fairly narrow feature 
around 9.3$\mum$ (as with FN Tau and V410 Anon 13), 
and a noisy downward slope to $\simali$11.4$\mum$, 
at which point the feature sharply drops.  
While large grains can provide flat-topped 10$\mum$ features 
of low contrast, they cannot produce the 9.3$\mum$ opacity
sharp peak; therefore, only a modest amount of large grains 
are included.  Between the local minima at 11.8 and 13.2$\mum$ 
in the emissivity, a small feature rises above 
the pixel-to-pixel noise, followed by a rise past 
13.5$\mum$ to longer wavelengths.  
Crystalline pyroxene can account for the 9.3$\mum$ peak, 
while the 11.4$\mum$ inflection indicates forsterite.  
As with V410 Anon 13, quartz is consistent within 
the noise near 12.5$\mum$.  
In addition, quartz provides, as with Hen 3-600A, 
a boost to the crystalline pyroxene at 9.3$\mum$ 
to increase its prominence, while at the same time 
giving rise to spectral features in the 12 to 13$\mum$ 
region accounting for some of the noisy features 
between 11.8 and 13.2$\mum$.  
Amorphous pyroxene is included to provide overall 
roundness to the 10$\mum$ feature.  
As with FN Tau, the dust model underestimates 
the derived CY Tau emissivity near 14$\mum$.

\section{Discussion}

\subsection{Lack of Processing of Silicates in Transitional Disks}

\citet{ld01} conclude that no more than 5\% of the Si 
in the diffuse ISM is in crystalline silicates.
\citet{kemp04} place an upper limit of $\simali$2.2\% 
for the silicate dust toward the Galactic center.
The result of $\simali$0.1\% mass fraction of crystalline 
silicates (forsterite, crystalline pyroxene, and quartz) 
for CoKu Tau/4 and nearly zero crystalline silicates 
for DM Tau and GM Aur (Table 4), is therefore consistent 
with nearly no processing of silicate dust in these disks, 
and with the material in these disks having originated 
from the ISM.  This seems to indicate that little processing 
occurs in the outer disks, or, if processing does occur in 
the inner disk, the dust in the inner disk is not transported 
efficiently to the outer disk.  
Perhaps the material in the inner optically thin disk 
of GM Aur was located between 10 and 24\,AU immediately 
before the planet formed near 24\,AU.  
This material would have experienced no processing, 
as in the case of CoKu Tau/4. Perhaps a planet opened 
up a disk gap at $\simali$24\,AU, halting inward 
accretion at 24\,AU of material from the outer disk.  
The disk inside of 24\,AU would continue accreting, 
and perhaps the last of this inner disk of nonprocessed 
dust is seen just before it gets accreted onto the star.  
In Figure 7 the mass fraction of crystalline silicates
$M_{\rm cryst.sil}$/$M_{\rm sil}$ versus wall radius 
is plotted for the five transitional disks.  
Note the lack of crystallinity for disks with walls beyond 2\,AU.

\subsection{Comparison of T Tauri Stars to Herbig Ae/Be Stars}

The spectra of the rest of the 1--3\,Myr old Taurus sample, 
all of which have inner disks more substantial than those 
of CoKu Tau/4, DM Tau, and GM Aur, indicate that 
crystalline silicates constitute at least $\simali$0.5\% 
(by mass) of all silicates in each disk.  
Whatever is responsible for producing the crystalline grains 
in T Tauri disks therefore appears linked to the inner 3\,AU 
disk region, a finding in line with the spectrointerferometric 
observations of Herbig Ae/Be disks \citep{vb04}.  
A substantial crystalline component is in place 
in the ``full'' T Tauri disks (defined in \S3.2) 
at 1--3\,Myr.  In our sample of TTSs, a majority of 
the silicate dust mass is always found in amorphous silicates.

The disks in the sample with lower mass fractions 
of crystalline silicates indicate forsterite 
but no crystalline pyroxene, while the disks 
with higher crystalline silicate mass fractions 
indicate both types of crystalline silicates.  
This is consistent with the trend reported by \citet{vb05} 
regarding the disks around Herbig Ae stars.  
Confirming this trend will require a larger sample 
and a more thorough investigation of the degeneracy 
between forsterite and crystalline pyroxene 
(see discussion in \S3.6).

All spectra except V410 Anon 13 indicate some 
large grains, but a greater crystalline mass 
fraction does not appear to correlate with 
a greater large-grain mass fraction (see Figure 8) 
for the T Tauri objects in our sample.  
See the discussion in \S3.5 on how grain sizes 
and shapes assumed in this study compare to those 
in other studies \citep{bouw01,vb05}.  
Because quartz accounts for less than $\simali$3\% 
of the total dust mass in the models, 
this implies that the mass fraction of forsterite
and crystalline pyroxene does not correlate with 
that of the 5$\mum$ sized porous amorphous silicate grains.  
It seems reasonable that, over time, the mass fraction 
of crystalline grains would increase, as should the mass 
fraction of large grains in disks around stars.  
However, crystallization and grain growth are not 
the same processes--crystallization involves 
changing the grain structure at the atomic level, 
while grain growth involves the gradual clumping
together of small grains to aggregate into large grains.  
The rates of increase of the crystalline grain mass fraction 
and of the large grain mass fraction should not necessarily 
be the same for even one single YSO disk.  
Furthermore, silicate grain growth and crystallization of 
silicate dust might proceed at different rates 
in disks with different accretion rates, mass, etc.  
It is not clear why large-grain mass fraction and 
crystalline grain mass fraction should be correlated 
in Herbig Ae/Be stars \citep{vb05} but not in T Tauri stars.

It is also curious that V410 Anon 13 indicates
a substantial mass fraction of crystalline silicates
but a negligible mass fraction of large grains.
If, as proposed by \citet{weid97}, large grains 
in disks settle more quickly than smaller grains 
toward the optically thick midplane, and vertical 
convection is negligible (as is expected due to 
the vertical temperature inversion in disk atmospheres),
then the lack of larger grains in a disk may simply be 
due to the fact that large grains have settled from the optically 
thin atmosphere of the disk, where they can be seen, down 
to the optically thick midplane.  A high mass fraction
of very large grains could be present in the V410 Anon 13 
disk atmosphere and only produce a featureless continuum.  

\Citet{vb05} report that Herbig Ae/Be stars of mass 
greater than 2.5$\msun$ have consistently high mass
fractions of crystalline silicates ($\sim$20\% to $\sim$30\%), 
and that Herbig Ae/Be stars less massive than 2.5$\msun$ 
have crystalline silicate mass fractions varying 
widely between 0 and $\simali$20\%. 
It is found that, like the lower mass Herbig Ae/Be stars, 
TTSs of mass $\simali$1.2$\msun$ down to $\simali$0.1 $\msun$ 
exhibit large dispersions in the mass fractions of
crystalline silicates.  Silicates in the disks around 
the lowest mass stars in our sample--FN Tau, V410 Anon 13, 
GG Tau B--have the highest crystalline percentages 
(between $\simali$27\% and $\simali$30\%) in our sample.  
According to the mass tracks from several different 
stellar evolutionary models, 
FN Tau \citep[spectral type M5;][]{kh95} is less massive, 
\citep[$\simali$0.28$\msun$;][]{sie00} than all other stars 
in our sample except V410 Anon 13 and the component stars 
of the GG Tau B pair.  The disks in our sample with the highest 
mass fractions of crystalline silicates surround 
the lowest mass stars.  A study of silicate crystallinity 
as a function of stellar mass for a much larger sample 
is underway (D. M. Watson et al. 2006, in preparation).  
The mass fraction of crystalline silicates
$M_{\rm cryst.sil}$/$M_{\rm sil}$ versus stellar 
mass is plotted in Figure 9.  
Using the mass fraction of only forsterite and crystalline 
pyroxene instead of $M_{\rm cryst.sil}$ in Figure 9 would 
not greatly change it as quartz is never a major 
component of the total crystalline grain mass.  
The emissivities of Hen 3-600 A and GG Tau B
(both known as binary systems), indicate some 
of the highest mass fractions of crystalline grains 
in our sample; the emissivity of the binary GG Tau A 
(which is coeval with GG Tau B, according to \citet{wgrs99}), 
however, indicates a low mass fraction of crystalline silicates.

\subsection{Production of Crystalline Silicate Dust}

Numerous mechanisms have been proposed for processing 
amorphous silicate grains into crystalline silicates.  
Based on the MIDI (Mid-IR Interferometric Instrument) spectra, 
\citet{vb04} argue that in the inner disk regions ($<$2\,AU), 
crystalline silicate grains are condensed directly from 
the gas phase.  Alternatively, solid amorphous silicates 
may be thermally annealed into crystalline grains; 
in thermal annealing, grains are heated sufficiently 
that constituent atoms redistribute themselves in
the grain in more energetically favored positions, 
forming a regular crystalline lattice \citep{bcmpb99}.  
Various mechanisms have been proposed to transport grains 
radially from the warm inner disk regions, where crystalline 
silicates may be produced, to the outer disk regions.  
These include convection and turbulent mixing \citep{boss04}, 
turbulent diffusion and large-scale circulation currents 
\citep{gail04}, and {\it X}-winds \citep{ssl96}.

Spectra of CoKu Tau/4 and DM Tau indicate that almost 
no crystalline silicates exist in the regions emitting
the mid-IR spectra. For CoKu Tau/4, this region is mostly 
the vertical wall at $\simali$10\,AU.  
For DM Tau, the region giving rise to much of its mid-IR 
spectrum is a combination of the wall at $\simali$3\,AU 
and the cooler optically thin atmosphere immediately 
beyond the wall \citep{cal05}. The lack of crystalline 
silicate grains in the outer disks of these two objects
(beyond $\simali$3\,AU for DM Tau and beyond $\simali$10\,AU 
for CoKu Tau/4) seems to suggest that very few crystalline 
silicate grains were radially transported to their outer disks; 
however, this may be expected in the case of planet formation 
and inner disk clearing.  After a planet forms, it can clear 
out a gap in a disk, and any newly formed crystalline silicate 
dust interior to the planet's orbit will either be accreted 
onto the planet or onto the star.  Furthermore, crystalline 
dust in the upper layers of the outer disk will accrete 
inward onto the newly formed planet.

Local production of crystalline grains is an alternative 
to radial transport.  Lightning has been proposed \citep{pil98} 
to heat dust grains and chondrules, but it is not clear exactly 
whether lightning in YSO disks might be generated \citep{dc02}.  
\citet{hd02} propose \textit{in situ} annealing at disk radii between 
$\simali$5 and 10\,AU, summarizing evidence that chondrules 
(which contain crystalline silicates) formed from circumstellar 
material in young disks between $\simali$5 and 10\,AU; 
therefore, they look to that region as the site of dust grain annealing.  
Harker \& Desch (2002) propose that submicron- to micron-sized 
dust grains are heated and annealed by thermal exchange 
with surrounding gas, itself heated by a disk shock front; 
the rate of cooling of the grains is then determined by 
the rate of cooling of the surrounding disk gas.  
Their simulations, using models described by 
Desch \& Connolly (2002), indicate that silicate grains 
can be annealed in 5\,km\,s$^{-1}$ shocks out to 10\,AU.  
In some simulations, 250$\mum$ radius chondrules 
were completely evaporated. 
According to \citet{bd05}, shock fronts have been seen 
in simulations of gravitationally unstable disks 
as transient phenomena.  To initiate gravitational instability, 
\citet{wood96} asserts that the ratio of the disk mass to the stellar mass 
must be around 0.3 to 0.5; however, in some numerical simulations, 
disk-to-star mass ratios as low as 0.05 \citep{bodu05} initiate 
the growth of gravitational instability in a disk, 
resulting in the formation of a spiral arm.  
This spiral arm would then be the site of shock annealing.

In an attempt to characterize the degree of likelihood 
of such shock processing to occur, disk mass relative to 
stellar mass in our small sample is considered.  
It is not known to which component of the GG Tau A binary 
the GG Tau A disk emission belongs, so it is not included 
in our discussion of the effect of the ratio of the disk mass to the stellar mass 
on the mass fraction of crystalline grains; 
for the same reason, GG Tau B is not included.  
In addition, there are no disk mass estimates 
for V410 Anon 13 and Hen 3-600 A, so they are not included.  
In Figure 10 we plot the disk-to-star mass ratio 
versus the mass fraction of crystalline silicates.  
There is no clear correlation between the disk-to-star mass 
ratio and the silicate crystallinity.  
As with Figures 8 and 9, when quartz is excluded 
from the total crystalline mass fraction, the conclusion 
does not change (i.e. there is no correlation between 
the disk-to-star mass ratio and the total mass of 
crystalline pyroxene and forsterite).

It is possible that planet formation may obscure any 
correlation between the disk-to-star mass ratio and 
the crystalline silicate mass fraction.  
Gravitational instabilities may induce disk shocks.  
If planets form by gravitational instability, 
which should be correlated with the disk-to-star mass ratio, 
then planets should form when disk shocks are in the process 
of annealing silicate grains.  Any crystalline silicates 
produced by disk shocks, which are interior to 
the forming planet, would then either accrete onto 
the planet or onto the star, thus eventually removing 
any evidence of them.  The relation (or lack thereof) 
of the disk mass-to-star mass ratio with the crystalline mass 
fraction for a much larger sample of objects in the Taurus-Auriga 
star-forming region will be studied in a forthcoming paper.

In Figure 11 is plotted for the full disks 
the mass fraction of crystalline silicates
versus $\beta_{9.9}$, the measure of the contrast 
of the 10$\mum$ feature to the underlying continuum.  
Note that the disks with lower crystallinity have 
higher $\beta_{9.9}$, and those with high degrees
of crystallinity have lower $\beta_{9.9}$.  
The mass fraction of large silicate grains 
versus $\beta_{9.9}$ is plotted in Figure 12.  
Here an upper envelope, 
decreasing for increasing $\beta_{9.9}$, 
to the large grain fraction may be present.  
Others have observed in Herbig Ae/Be stars \citep{vb03} 
and T Tauri stars \citep{prz03,ks05,ks06} 
a trend of decreasing the 10$\mum$ feature contrast 
with increased dust processing 
(grain growth and/or dust crystallization).  
The decrease of both the crystallinity and 
the large grains mass fractions with 
increasing $\beta_{9.9}$ in our sample of 12 T Tauri disks 
is consistent with the earlier studies.

At 10\,Myr, Hen 3-600 A presents a disk with a large mass 
fraction of crystalline silicates, while TW Hya does not. 
TW Hya ($\simali$0.7$\msun$; see Webb et al.\ 1999
and Siess et al.\ 2000) is hypothesized to have 
a protoplanetary or planetary companion 
orbiting between $\simali$2 and 3\,AU \citep{cal02} 
and is reported to have a fairly massive disk of
$\sim$0.03 $\msun$ \citep{wil00}.  
\citet{uch04} show that the optically thin disk 
inside of $\simali$3.3\,AU is responsible for 
most of the 10$\mum$ emission in TW Hya.  
While the spectrum of TW Hya does indicate 
a small amount of forsterite and quartz, 
and a substantial fraction of large grains, 
it does not indicate crystalline pyroxene; 
the derived crystalline mass fractions for TW Hya 
are similar to those of GG Tau A and IP Tau, 
but TW Hya is much older than those stars.  
Perhaps the protoplanetary companion to TW Hya 
orbiting at 2-3\,AU prevents outward radial 
transfer of crystalline silicate grains produced 
in the inner disk; alternatively, perhaps the small 
star-planet separation allows only a small zone 
in which shocks can form, annealing fewer silicate grains.  
The transition from optically thin inner disk to optically 
thick outer disk is modeled to be at $\simali$1.3\,AU 
in Hen 3-600 A \citep{uch04}, indicating a more substantial 
disk in Hen 3-600 A than in TW Hya.  
Hen 3-600 A is the only one transitional disk with a substantial
degree of crystallinity (see Figure 7).  
The denser disk between 1.3 and 3.3\,AU in Hen 3-600 A 
might lead to either more inner disk crystalline silicate 
production and freer radial transport of 
crystalline grains or more substantial shock 
processing than in TW Hya.

\subsection{Possible Connection between Quartz and Amorphous Pyroxene}

There is an indication of increasing mass fraction 
of quartz with decreasing mass fraction ratio of 
amorphous olivine 
to amorphous pyroxene 
(Figure 13).  The IRS spectra of TW Hya, Hen 3-600 A, 
and GG Tau A all indicate the presence of quartz and 
low amorphous olivine to amorphous pyroxene mass ratios.  
We caution, however, that our sample size is small (12 spectra), 
and that further study of this possible trend, using a much 
larger sample of spectra, will be needed to verify this finding.  
Furthermore, some of the mass in quartz in these disks 
could be in forms of silica other than $\alpha$ quartz, 
which would affect this trend.  
These issues will be explored in a future paper.

\section{Summary and Conclusions}

The 10$\mum$ features in 10 TTS spectra from the Taurus-Auriga 
star-forming region and 2 spectra from the TW Hydrae Association 
are modeled by deriving the emissivities from IRS spectra of 
the optically thin dust emission.  Through an iterative process 
of varying dust grain mass fractions of various dust species 
to match the dust cross-sections to the derived emissivities 
and using the derived 
$\sigma$($\sim$20 $\mu$m)/$\sigma$($\sim$10 $\mu$m) ratios 
to recompute the dust temperature 
(and therefore self-consistently define the emissivity), 
the modeled emissivities are fit with a multi-component dust model.

To test the simple single-temperature approach 
of deriving and fitting the 10$\mum$ feature emissivities,
a wavelength-dependent dust cross-section similar to the one generated from 
the fit to IP Tau was used to determine the emergent spectrum
in a model of its emission, and it was found that the fit to 
the spectrum of IP Tau requires a significant source of 
absorption at visible/near-IR wavelengths.  
To fit the interstellar extinction, \citet{dl84} also required
a source of absorption at these wavelengths in addition to 
that from known silicates.  Consequently, graphite and 
``astronomical silicates'' were used to determine 
the radial and vertical temperature structure of 
the model disk for IP Tau, and a dust mixture similar to the best-fit dust mixture
for this object was used to compute the emergent spectrum 
from the model disk.  This full radiative transfer model 
gave an excellent fit to the 5--12$\mum$ spectrum of IP Tau, 
supporting the simplified emissivity modeling technique used here.  
The model predicts too much emission at wavelengths longer 
than 12$\mum$, indicating too much emission in the model 
from cooler dust grains, suggesting there is settling of 
dust from the disk atmosphere to the midplane.

Our 12 object sample indicates the following:
\begin{itemize}
\vspace{-4mm}
\item Almost all the T Tauri disks show some degree of silicate 
      crystallinity, irrespective of star mass or age.
\item Transitional disks, disks whose inner portions are either 
      partially or nearly totally cleared of small dust grains, 
      usually indicate very few crystalline silicate grains.
\item Theories of radial transport of crystalline silicates 
      in T Tauri disks are constrained to explain the ISM-like 
      lack of crystalline silicates outward of 3\,AU in DM Tau 
      and outward of 10\,AU in CoKu Tau/4, both of which are 
      1--3 Myr old systems.
\item No obvious correlation of the mass fraction of crystalline 
      silicate grains with that of large silicate grains exists 
      for the T Tauri sample.  This contrasts with the results 
      of \citet{vb05} for Herbig Ae/Be systems 
      (see discussion in \S3.5 regarding differences in grain shape 
      and size between this study and Bouwman et al. [2001] and van Boekel et al. [2005]).
\item The 10$\mum$ features always indicate a majority of mass 
      in amorphous silicate grains.
\item Crystalline pyroxene is usually accompanied by forsterite, 
      but the reverse is not necessarily true.
\item Very low mass stars can have relatively large amounts of 
      crystalline silicates in their surrounding disks.
\item No clear trend exists between the mass fraction of
      crystalline silicates and the disk-to-star mass ratio.
\item For full disks, high contrast of the 10$\mum$ feature, 
      as measured by $\beta_{9.9}$, indicates both low crystallinity 
      and small mass fraction of large grains; 
      decreasing contrast of the 10$\mum$ feature 
      for these disks indicates increasing crystallinity 
      and a range of mass fractions of large grains.
      This is consistent with studies of Herbig Ae/Be stars 
      \citep{vb03} and previous studies of T Tauri stars 
      \citep{prz03,ks05,ks06}.
\item There is an indication that higher quartz mass fraction 
      accompanies lower amorphous olivine to amorphous pyroxene ratio.
\end{itemize}
      
One of the surprising results of this study 
is the lack of correlation of the silicate crystallinity 
with anything else.  This may reflect the diversity in 
the evolutionary histories of disks (e.g., some make more 
crystalline silicates than others), which may be 
a consequence of initial conditions 
(e.g., initial disk mass and angular momentum) 
and/or environment (e.g., binarity).  
Considerable diversity may result from 
the process of giant planet formation.  
For example, suppose that within the first 1--3\,Myr 
of existence, all disks accumulate crystalline silicate grains 
in their inner disks either by inner disk heating and outward 
radial transport or by local annealing of silicate grains.  
Suppose further that a giant planet forms in the inner disk.  
According to \citet{quil04}, the planet clears a gap in 
the disk and absorbs the accretion from the outer disk.  
The inner disk is cleared by accretion onto the star
in $\simali$10$^{5}$ yr, leaving a transitional disk 
with the characteristics of the disk around either 
CoKu Tau/4 or DM Tau.  The remnant outer disk has 
very few crystalline silicate dust grains remaining, 
also like CoKu Tau/4.  If the disk is sufficiently massive, 
as is the case with DM Tau, the planet can migrate radially inward.  
If this planet formation and migration scenario happens
with sufficient frequency in T Tauri disks, 
a large number of T Tauri disks with near IR excesses
not indicating transitional disks but with 10$\mum$ features 
indicating low crystallinity, like FM Tau, will be observed.  
Both giant planet formation and crystallization can depend on 
the disk/star mass ratio, but crystallized dust can be removed 
after the planet forms.

In a future paper, the relative abundances of various dust species 
in a larger sample of T Tauri disks will be quantified, 
and the abundances of dust species will be compared to 
other dust species and also to various stellar properties.  
Another goal is to incorporate the best-fit dust cross-sections 
for our T Tauri objects into the sophisticated radiative transfer
models described by \citet{cal91,cal92}, 
and \citet{daless98,daless99,daless01} 
to model these objects' SEDs and spectra 
self-consistently over much wider wavelength ranges.

\acknowledgments This work is based on observations made with 
the {\it Spitzer Space Telescope}, which is operated by 
the Jet Propulsion Laboratory, California Institute of Technology 
under NASA contract 1407.  Support for this work was provided by 
NASA through contract 1257184 issued by JPL/Caltech and 
through the Spitzer Fellowship Program, under award 011 808-001, 
and JPL contract 960803 to Cornell University, 
and Cornell subcontracts 31419-5714 to the University of Rochester.  
The authors thank Ciska Markwick-Kemper for offering 10$\mum$ data 
for GCS\,3 and for helpful comments on the paper. 
The authors also acknowledge helpful comments from Luke Keller.  
P. D. acknowledges grants from PAPIIT, DGAPA, UNAM and CONACyT, M\'{e}xico.  
N. C. and L. H. acknowledge support from NASA grant NAG5-13210, 
STScI grant AR-09524.01-A, and NASA Origins grant NAG5-9670.  
A.L. acknowledges support from the University of Missouri Summer 
Research Fellowship, the University of Missouri Research Board, 
and the NASA award P20436.
SMART was developed by the IRS Team at Cornell University 
and is available through the {\it Spitzer} Science Center at Caltech.  
This publication makes use of the Jena-St. Petersburg Database 
of Optical Constants \citep{hen99}.

\appendix

\section*{Appendix}

In our attempts to fit the dust emission feature 
at $\simali$11.3$\mum$ in many of our emissivities, 
we unexpectedly found a very close resemblance 
between the opacity profile of forsterite grains 
with a CDE shape distribution and that of 
porous forsterite grains with a volume fraction of 
vacuum $f$\,=\,0.6 (i.e. 60\% of the grain volume is
vacuum) and a fractal dimension of $D$\,=\,2.766.
For these porous grains, we used the Bruggeman effective 
medium theory (Bohren \& Huffman 1983) and the dielectric 
constants of forsterite of \citet{fab01} to generate 
the effective dielectric constants for each crystalline axis.  
For a given crystalline axis, we generate opacity from 
Mie theory using the effective dielectric constants 
for that axis, assuming a spherical grain radius of 
0.1$\mum$.  We then obtain the opacity for 
the porous forsterite grain by averaging the opacities 
of all three crystalline axes.  The resulting porous 
forsterite opacity is nearly identical to that derived 
using the same set of dielectric constants and 
assuming a CDE shape distribution (Figure 14).  
We made the same comparison using parameters 
for the ordinary and extraordinary rays of 
$\alpha$ quartz (Wenrich \& Christensen 1996).  
As for forsterite, we used $f$\,=\,0.6, $D$\,=\,2.766, 
and grain radius of 0.1$\mum$; 
however, for quartz we add the sum of \twothirds of
the opacity generated from the ordinary ray 
and \onethird of that from the extraordinary ray.  
As for forsterite, the quartz opacities generated 
assuming porous grains and grains with a CDE distribution 
are nearly identical (see Figure 14).
This suggests that the near-equivalence of
the Bruggeman EMT plus Mie theory to CDE is 
mathematical in nature and not specific to
the optical properties of the dust material in question.  

\clearpage

\begin{figure}[t] 
  \epsscale{0.6}
  \plotone{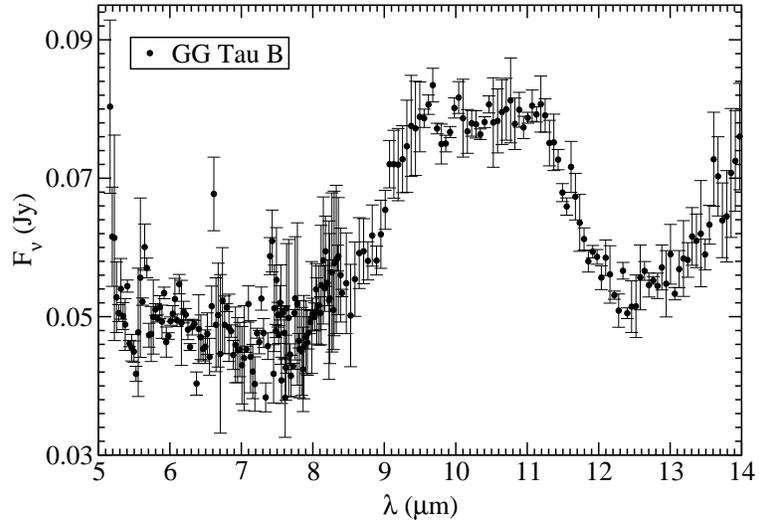}
  \caption{The 5--14$\mum$ IRS spectrum of GG Tau B.}
\end{figure}


\begin{figure}[t] 
  \epsscale{0.6}
  \plotone{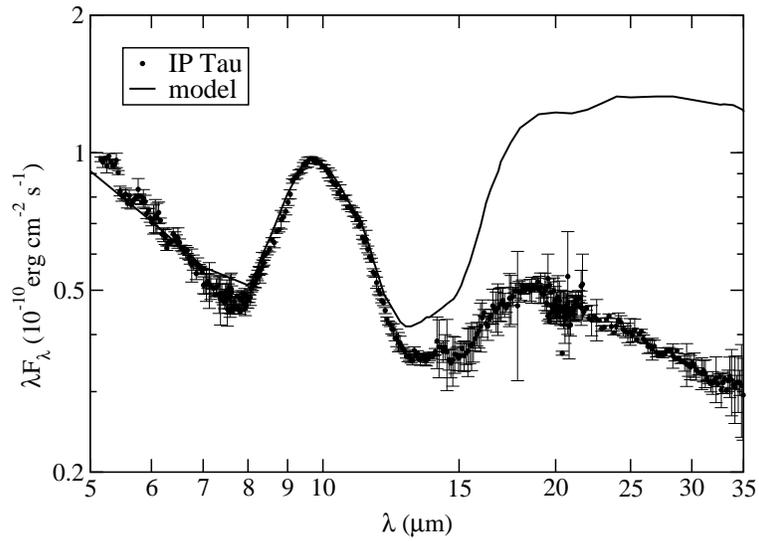}
  \caption{The 5--35$\mum$ IRS spectrum of IP Tau 
           fit by a disk model using a dust mixture similar to that from Table 4 for IP Tau.}
\end{figure}

\clearpage

\begin{figure}[t] 
  \epsscale{0.6}
  \plotone{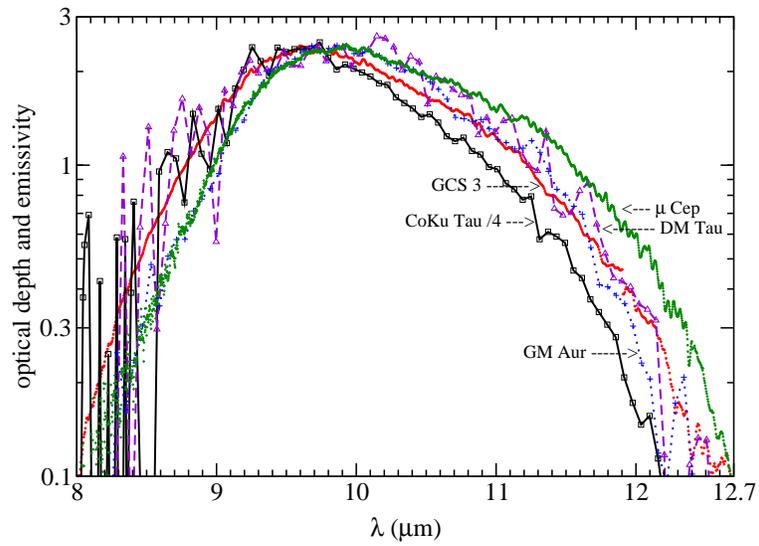}
  \caption{The 10$\mum$ silicate profiles of GCS 3, 
           CoKu Tau/4, DM Tau, GM Aur, and $\mu$ Cep.  
           The GCS\,3 profile of Kemper et al.\ (2004)
           is the interstellar silicate absorption 
           toward the Galactic center 
           and the other four profiles are scaled emissivities 
           with baselines subtracted (see Table 3).}
\end{figure}

\clearpage

\begin{figure}[t] 
  \vspace{-0.15in}
  \epsscale{0.9}
  \plotone{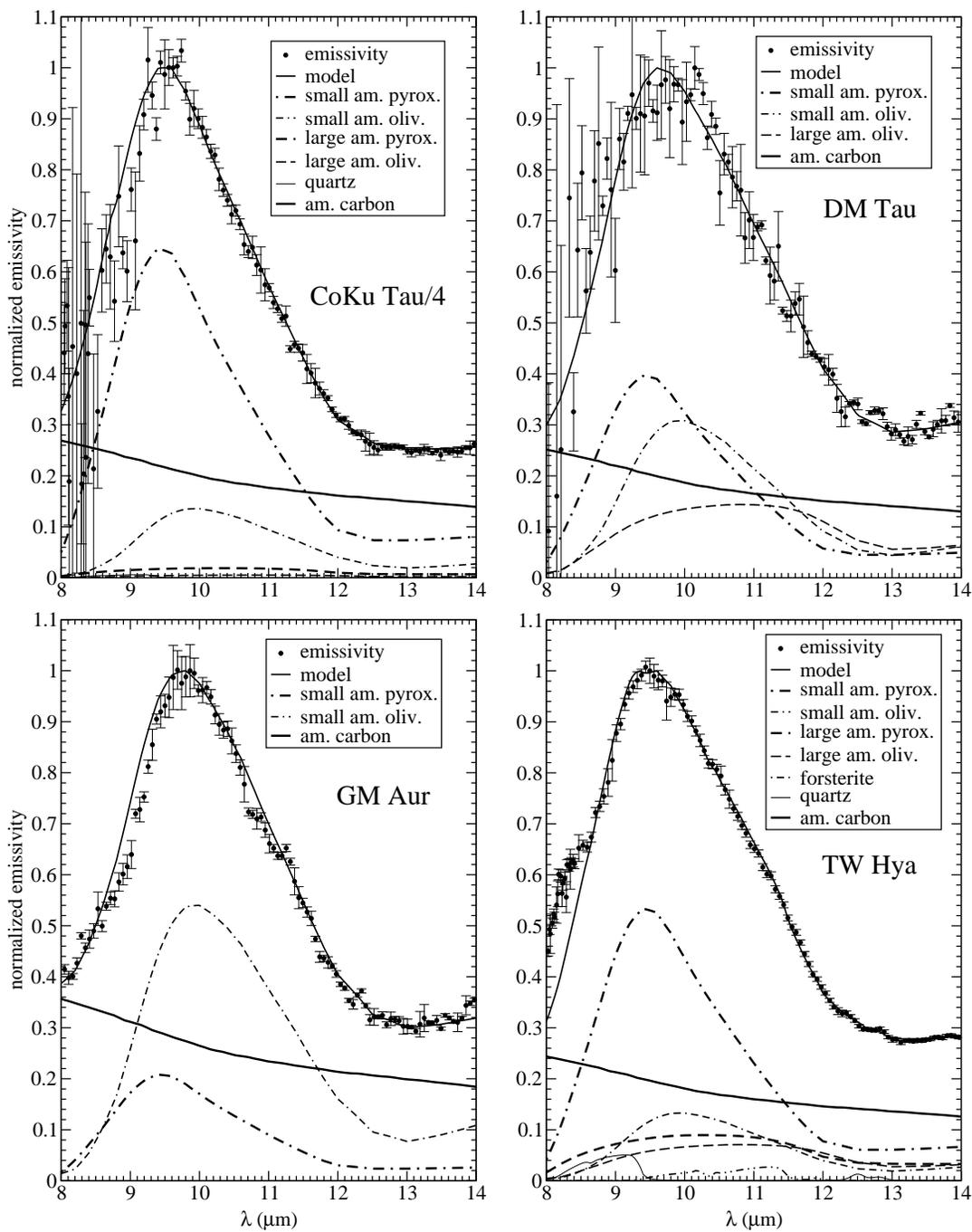}
  \vspace{-0.55in}
  \caption{Derived emissivities of CoKu Tau/4, DM Tau, 
           GM Aur, and TW Hya. Also displayed are the dust
           model spectra, as well as the spectrum of each 
           dust component.}
\end{figure}

\clearpage

\begin{figure}[t] 
  \epsscale{0.9}
  \plotone{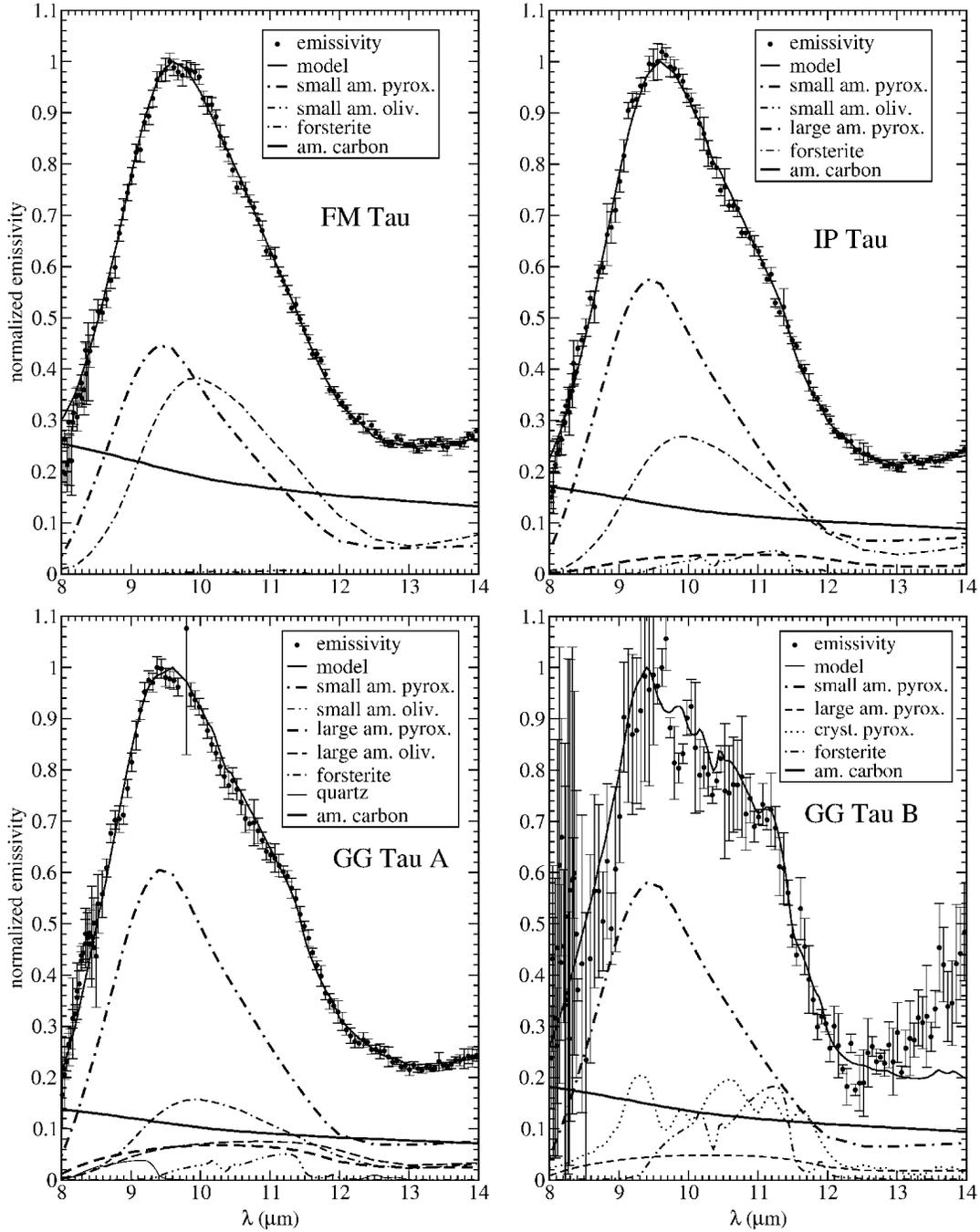}
  \vspace{-0.57in}
  \caption{Same as Fig. 4, but for FM Tau, IP Tau, GG Tau A, and GG Tau B.}
\end{figure}

\clearpage

\begin{figure}[t] 
  \epsscale{0.9}
  \plotone{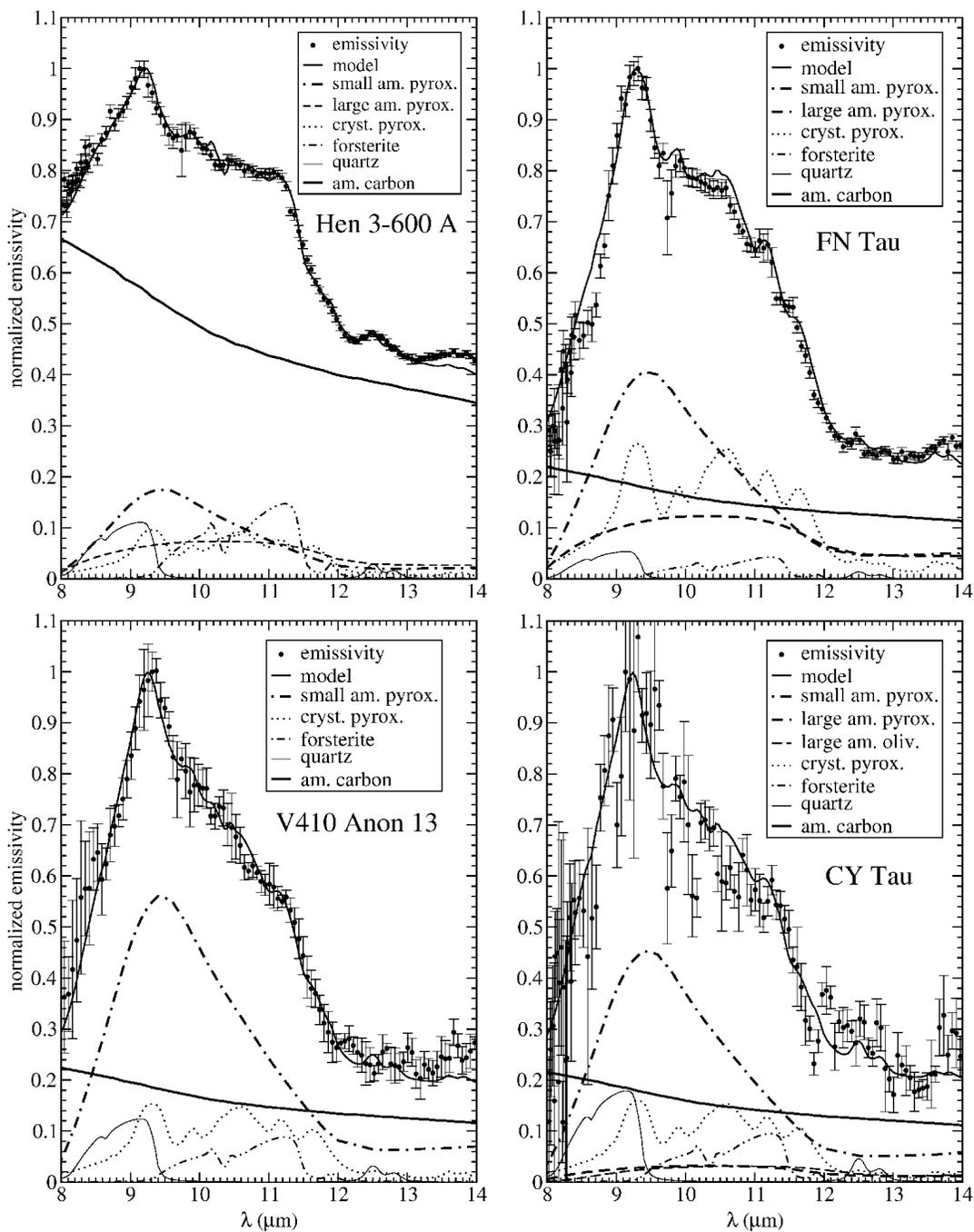}
  \vspace{-0.57in}
  \caption{Same as Fig. 4, but for 
           Hen 3-600 A, FN Tau, V410 Anon 13, and CY Tau.}
\end{figure}


\begin{figure}[t] 
  \epsscale{0.6}
  \plotone{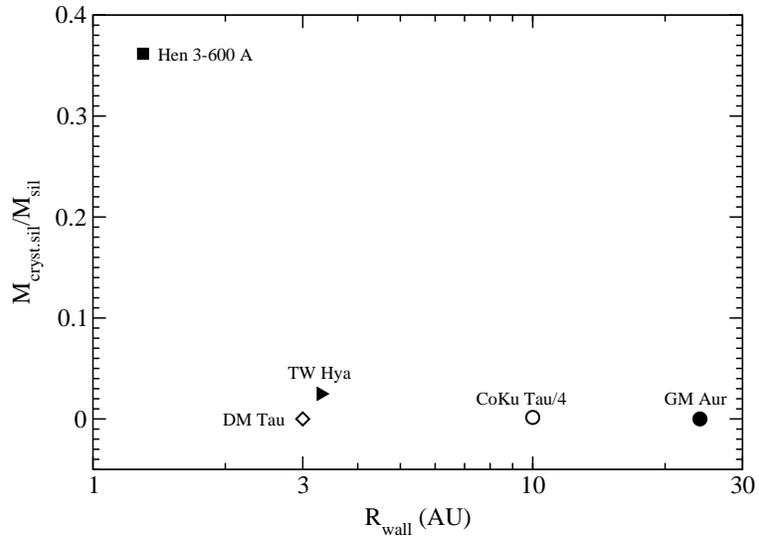}
  \caption{Mass fraction of crystalline silicate dust 
           vs. radial location of wall for transitional disks.}
\end{figure}


\begin{figure}[t] 
  \epsscale{0.6}
  \plotone{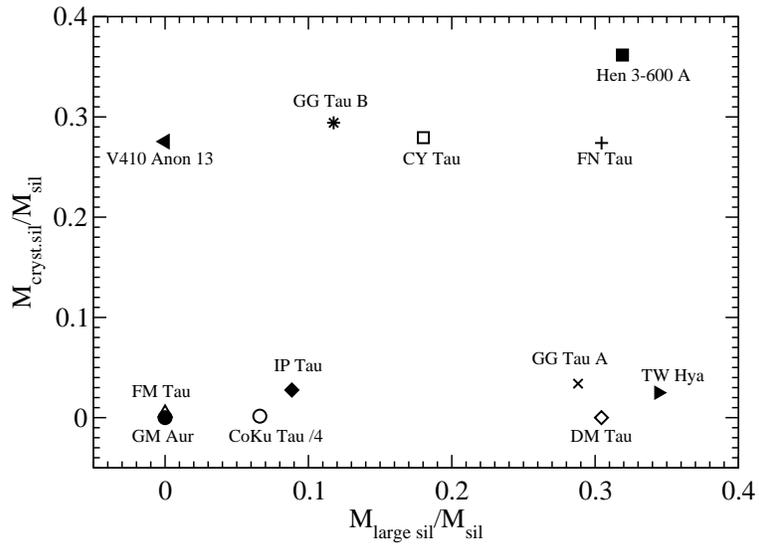}
  \caption{Mass fraction of crystalline silicates vs. that of
           large silicate grains for all 12 T Tauri stars.}
\end{figure}


\begin{figure}[t] 
  \epsscale{0.6}
  \plotone{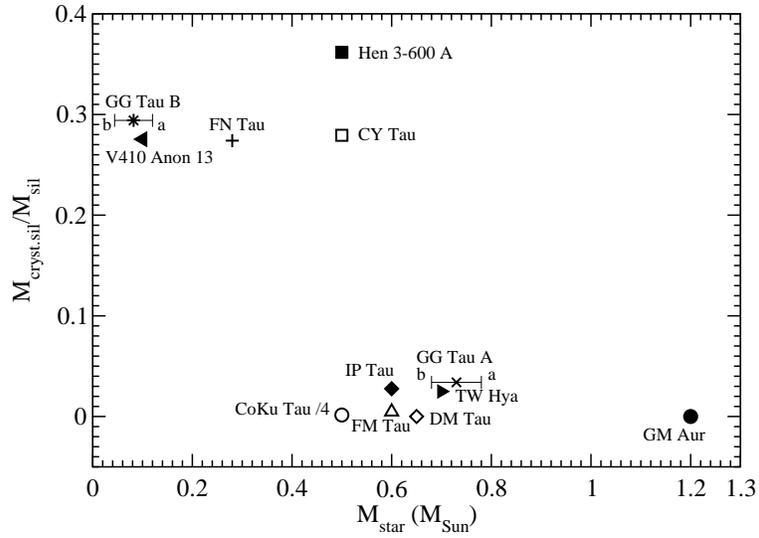}
  \caption{Mass fraction of crytalline silicate grains 
           vs. stellar mass for all 12 T Tauri stars.  
           The stellar mass plotted for Hen 3-600 A 
           (a spectroscopic binary) is the sum of 
           the masses of its two components.}
\end{figure}


\begin{figure}[t] 
  \epsscale{0.6}
  \plotone{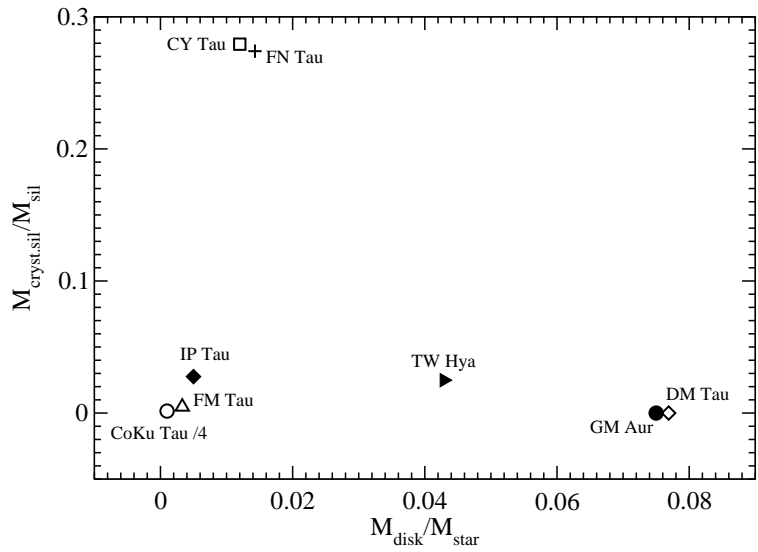}
  \caption{Mass fraction of crystalline silicate dust 
           vs. disk-to-star mass ratio.}
\end{figure}


\begin{figure}[t] 
  \epsscale{0.6}
  \plotone{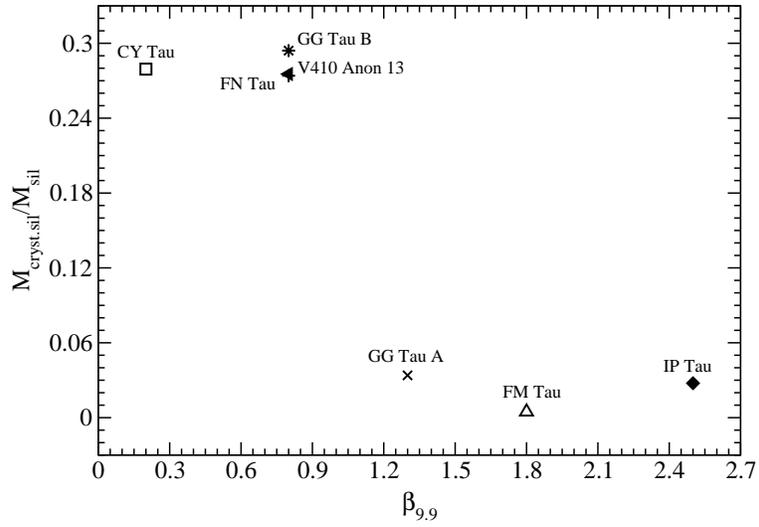}
  \caption{Mass fraction of crystalline silicate dust 
           vs. $\beta_{9.9}$ -- the continuum-subtracted 
           residual flux at 9.9$\mum$ to the 9.9$\mum$ continuum
           for ``full disks''.}
\end{figure}


\begin{figure}[t] 
  \epsscale{0.6}
  \plotone{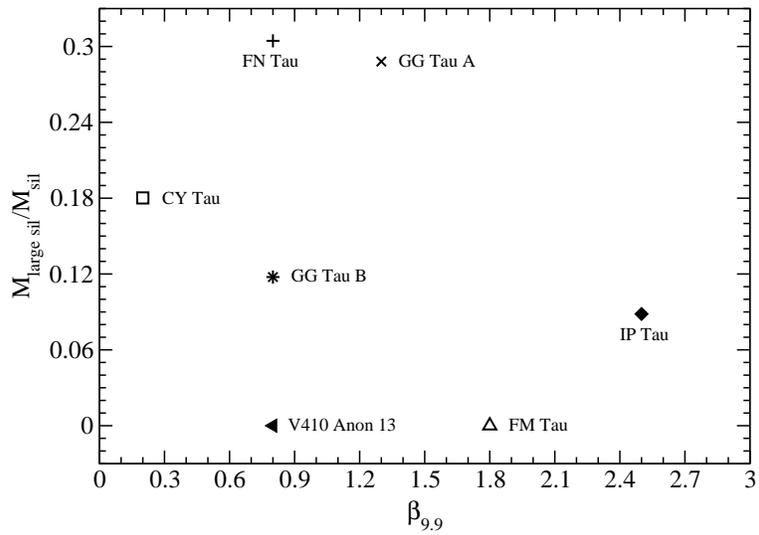}
  \caption{Mass fraction of large silicate grains 
           vs. $\beta_{9.9}$ for ``full disks.''}
\end{figure}

\clearpage

\begin{figure}[t] 
  \epsscale{0.6}
  \plotone{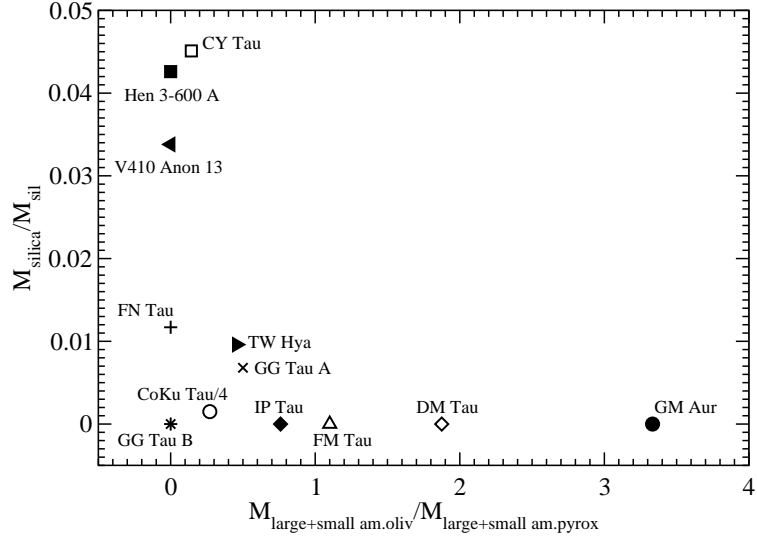}
  \caption{Mass fraction of quartz vs. mass ratio
           of amorphous olivine to amorphous pyroxene for 
           all 12 T Tauri stars.}
\end{figure}


\begin{figure}[t] 
  \epsscale{0.6}
  \plotone{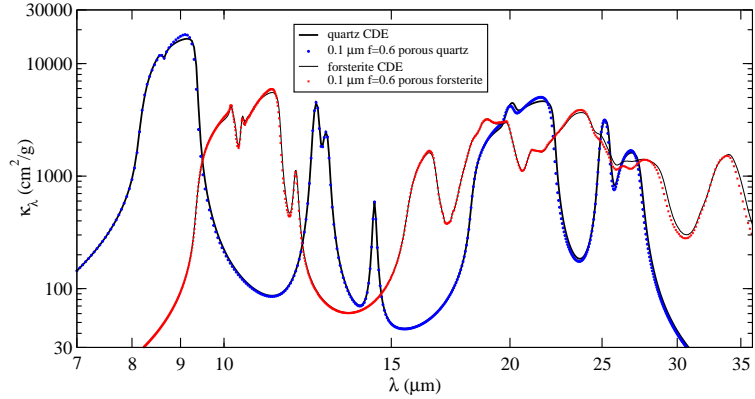}
  \caption{Comparison of opacity profile of solid 
           $\alpha$ quartz ({\it thick line}) and forsterite ({\it thin line})
           with a CDE shape distribution with that for porous 
           $\alpha$ quartz ({\it blue circles}) and forsterite ({\it red circles}).}
\end{figure}

\clearpage

\begin{deluxetable}{cccccccc}
\tablecaption{The T Tauri Sample \label{table1}}
\tablehead{
\colhead{} & \colhead{Spectral} &\colhead{$A_V$} 
           &\colhead{$T_{\rm eff}$} &\colhead{$L_*$} 
           &\colhead{$M_*$} &\colhead{$M_{\rm disk}$} 
           &\colhead{}\\
\colhead{Object}  & \colhead{Type} &\colhead{(mag)} 
           &\colhead{(K)} &\colhead{($\lsun$)} 
           &\colhead{($\msun$)} &\colhead{($10^{-3}\msun$)} 
           &\colhead{$M_{\rm disk}$/$M_*$}} 
\startdata
CoKu Tau/4 & M1.5$^{a}$ & 3.0$^{a}$ & 3720$^{a}$ & 0.61$^{a}$ & 0.5$^{j}$ & 0.5$^{k}$ & 0.001 \\
DM Tau & M1$^{b}$ & 0.5$^{b}$ & 3720$^{b}$ & 0.25 & 0.65$^{b}$ & 50$^{b}$ & 0.077 \\
GM Aur & K5$^{b}$ & 1.2$^{b}$ & 4730$^{b}$ & 0.83 & 1.2$^{b}$ & 90$^{b}$ & 0.075 \\
TW Hya & K7$^{c}$ & 0$^{d}$ & 4009$^{c}$ & 0.25$^{c}$ & 0.7$^{j}$ & 30$^{e}$ & 0.043 \\
FM Tau & M0 & 0.69 & 3850 & 0.32 & 0.6$^{j}$ & 2$^{k}$ & 0.003 \\ 
IP Tau & M0 & 0.24 & 3850 & 0.43 & 0.6$^{j}$ & 3$^{k}$ & 0.005 \\
GG Tau Aa & K7$^{f}$ & 0.8$^{f}$ & 4060$^{g}$ & 0.84$^{f}$ & 0.78$^{f}$ & \nodata  & \nodata \\
GG Tau Ab & M0.5$^{f}$ & 3.2$^{f}$ & 3850$^{g}$ & 0.71$^{f}$ & 0.68$^{f}$ & \nodata & \nodata \\
GG Tau Ba & M5$^{f}$ & 0.55$^{f}$ & 3050$^{f}$ & 0.08$^{f}$ & 0.12$^{f}$ & \nodata & \nodata \\
GG Tau Bb & M7$^{f}$ & 0$^{f}$ & 2820$^{f}$ & 0.02$^{f}$ & 0.044$^{f}$ & \nodata & \nodata \\
Hen 3-600 A & M3e$^{c}$ & 0.7$^{i}$ & 3350$^{c}$ & 0.2$^{c}$ & 0.5$^{j}$ & \nodata & \nodata \\
FN Tau & M5 & 1.35 & 3240 & 0.5 & 0.28$^{j}$ & 4$^{l}$ & 0.014 \\
V410 Anon 13 & M5.75$^{h}$ & 5.8$^{h}$ & 3000$^{h}$ & 0.077$^{h}$ & 0.1$^{h}$ & \nodata & \nodata \\
CY Tau & M1 & 0.1 & 3720 & 0.47 & 0.5$^{j}$ & 6$^{k}$ & 0.012 \\
\enddata
\tablecomments{\footnotesize For Hen 3-600 A, we determine the mass 
               for each of the two components of the spectroscopic 
               binary, assuming that (1) both components have 
               the same luminosity, (2) the two components' combined 
               luminosity is the same as that plotted in Figure 3 
               of \citet{webb99} for Hen 3-600 A, and
               (3) the effective temperatures of both components 
               are the same and equal to that plotted for Hen 3-600 A 
               in Figure 3 of \citet{webb99}.
               We then derive mass for one of the components 
               from \citet{sie00}, and add the masses 
               (to plot a point for Hen 3-600 A in Figure 7 and
               Table 1).}
\vspace{-3mm}
\tablecomments{\footnotesize All data from \citet{kh95} 
               unless otherwise noted.}
\tablenotetext{a}{\footnotesize from \citet{daless05}}
\tablenotetext{b}{\footnotesize from \citet{cal05}}
\tablenotetext{c}{\footnotesize from \citet{webb99}}
\tablenotetext{d}{\footnotesize from \citet{her04}}
\tablenotetext{e}{\footnotesize from \citet{wil00}}
\tablenotetext{f}{\footnotesize from \citet{wgrs99}}
\tablenotetext{g}{\footnotesize $T_{\rm eff}$ from \citet{kh95} 
                  based on the spectral type of \citet{wgrs99}}
\tablenotetext{h}{\footnotesize from \citet{fur05a}}
\vspace{-2mm}
\tablenotetext{i}{\footnotesize from \citet{gm01}}
\vspace{-2mm}
\tablenotetext{j}{\footnotesize using \citet{sie00}}
\vspace{-2mm}
\tablenotetext{k}{\footnotesize from \citet{aw05}}
\tablenotetext{l}{\footnotesize from \citet{bscg90}}
\end{deluxetable}

\clearpage

\begin{table}[h,t]
{\tiny
\caption[]{Emissivity Modeling Parameters \label{table2}}
\begin{tabular}{lcccccccccc}
\hline \hline
          & $\Omega_*$
          & T$_*$ 
          & F$_{10}$
          & power
          & $\lambda_s$ 
          & $\lambda_l$
          &  
          & T$_d$
          & $\Omega_{d}\tau_{max}$
          & \\
          Object$^{a}$
          & ($10^{-19}$\,sr)
          & (K)
          & (Jy)
          & index $m$
          & ($\mu$m)
          & ($\mu$m)
          & $\epsilon_l$/$\epsilon_s$
          & (K)
          & ($10^{-16}$\,sr)
          & $\beta_{9.9}^{b}$\\
\hline
$\mu$ Cep & 43300 & 3500 & \nodata & \nodata & 10.0 & 18.2 & 0.64 & 423 & 12650 & 5.1 \\
CoKu Tau/4 & 2.6 & 3720 & \nodata & \nodata & 10.0 & 20.0 & 0.50 & 121 & 1227 & * \\
DM Tau & 1.2 & 3720 & \nodata & \nodata & 9.5 & 19.0 & 0.59 & 160 & 89.4 & * \\
GM Aur & 1.5 & 4730 & \nodata & \nodata & 9.4 & 18.8 & 0.58 & 310 & 6.1 & * \\
TW Hya & 7.9 & 4009 & \nodata & \nodata & 9.65 & 19.3 & 0.56 & 193 & 293.8 & * \\
FM Tau & \nodata & \nodata & 0.131 & -0.388 & 10.0 & 20.0 & 0.50 & 222 & 41.4 & 1.8 \\
IP Tau & \nodata & \nodata & 0.090 & -0.936 & 10.0 & 20.0 & 0.52 & 259 & 15.5 & 2.5 \\
GG Tau A & \nodata & \nodata & 0.445 & -0.456 & 9.65 & 19.3 & 0.56 & 252 & 49.8 & 1.3 \\
GG Tau B & \nodata & \nodata & 0.042 & -0.370 & 10.0 & 20.0 & 0.67 & 252 & 3.3 & 0.8 \\
Hen 3-600 A & 19.3 & 3350 & \nodata & \nodata & 9.65 & 19.3 & 0.65 & 229 & 126.0 & * \\
FN Tau & \nodata & \nodata & 0.339 & 0.161 & 10.0 & 20.0 & 0.71 & 208 & 86.1 & 0.8 \\
V410 Anon 13 & \nodata & \nodata & 0.017 & -0.433 & 10.0 & 20.0 & 0.63 & 256 & 1.3 & 0.8 \\
CY Tau & \nodata & \nodata & 0.120 & -0.627 & 10.0 & 20.0 & 0.69 & 239 & 3.7 & 0.2 \\
\hline
\end{tabular}
\tablenotetext{a}{For the transitional disk sources 
               (CoKu Tau/4, DM Tau, GM Aur, TW Hya, Hen 3-600 A), 
               we subtract $\Omega_{*}B_{\nu}(T_{*})$, 
               representing the photospheric emission, from
               the dereddened IRS spectra to isolate 
               the optically thin emission in the 10$\mum$ complex; 
               for other sources, a power law is fit to
               the $\lambda$\,$<$\,8$\mum$ IRS spectrum and
               subtracted.  The power law is represented by 
               $F_{\nu}(\lambda)$\,=\,$F_{10}\,(\lambda/10\mum)^{m}$, 
               where F$_{10}$ is the flux density at 10$\mum$ in Janskys.  
               To determine the dust temperature $T_d$, 
               the excess flux and the model opacity are 
               determined at wavelength regions centered 
               at $\lambda_{\rm short}$ and $\lambda_{\rm long}$.  
               Dividing these blackbody- or continuum-subtracted 
               residuals by a Planck function of $T_d$, 
               gives the relative emissivities.  
               These emissivities are divided by 
               the normalization constant $\Omega_{d}\tau_{max}$, where $\tau_{max}$ is the maximum optical depth in the 10 $\mu$m feature of the optically thin residual flux.}
\tablenotetext{b}{\footnotesize $\beta_{9.9}$ is the ratio of
                  the continuum-subtracted residual 
                  at $\simali$9.9$\mum$ to the $\simali$9.9$\mum$ 
                  continuum (which was subtracted to derive the residuals).  
                  The asterisks (*) for CoKu Tau/4, DM Tau, GM Aur, TW Hya, and Hen 3-600 A indicate 
                  that there is no detectable disk flux continuum
                  underneath their 10$\mum$ feature, 
                  so their $\beta_{9.9}$ are effectively infinity.  
                  For $\mu$ Cep, the continuum is the stellar blackbody.}
}
\end{table}

\clearpage

\begin{deluxetable}{cccc}
\tablecaption{Parameters for Silicate Profiles\label{table3}}
\tablehead{\colhead{Object} &\colhead{$\epsilon_{8}$} 
                            &\colhead{$\epsilon_{13}$}
                            &\colhead{scaling factor}}
\startdata
$\mu$ Cep & 0.21 & 0.44 & 3.4 \\
CoKu Tau /4 & 0.33 & 0.25 & 3.4 \\
DM Tau & 0.53 & 0.29 & 4.5 \\
GM Aur & 0.42 & 0.30 & 3.9 \\
\enddata

\tablecomments{The 10$\mum$ profiles plotted in Figure 4
               were derived by subtracting a linear 
               baseline from the emissivities of CoKu Tau/4, 
               DM Tau, and GM Aur from Figure 4 
               and $\mu$ Cep (not shown here),
               and scaling the residuals to 
               the peak value of the optical depth of GCS 3. 
               The second and third columns give the values 
               of the emissivity baseline at $\simali$8
               and $\simali$13$\mum$, respectively.   
               After subtracting the baseline, 
               the residual emissivity was multiplied 
               by the fourth column.}
\end{deluxetable}

\clearpage

\begin{table}[h,t]
{\tiny
\caption[]{Dust Mass Percentages and Reduced $\chi^2$ \label{table4}}
\begin{tabular}{lccccccccccl}
\hline \hline
          & small
          & small 
          & large
          & large
          & am. 
          & cryst.
          & cryst. 
          & cryst.
          &
          & cryst.
          & large\\
          Object 
          & am. pyrox\tablenotemark{a}
          & am. oliv\tablenotemark{b}
          & am. pyrox\tablenotemark{c}
          & am. oliv\tablenotemark{d}
          & carbon\tablenotemark{e}
          & pyrox\tablenotemark{f}
          & forst\tablenotemark{g}
          & quartz\tablenotemark{h}
          & $\frac{\chi^2}{\rm d.o.f.}$\tablenotemark{i}
          & silicates\tablenotemark{j}
          & silicates\tablenotemark{h}\\
\hline
$\mu$ Cep & 0 & 100 & 0 & 0 & 0 & 0 & 0 & 0 & \nodata & 0 & 0\\
CoKu Tau/4 & 47.8 & 12.9 & 3.3 & 1.0 & 34.9 & 0 & 0 & 0.1 & 2.3 & 0.15 & 6.6\\
DM Tau & 25.1 & 25.1 & 0 & 21.9 & 27.9 & 0 & 0 & 0 & 6.6 & 0 & 30.4\\
GM Aur & 13.6 & 45.5 & 0 & 0 & 40.9 & 0 & 0 & 0 & 9.8 & 0 & 0\\
TW Hya & 34.6 & 11.1 & 13.8 & 11.1 & 27.7 & 0 & 1.1 & 0.7 & 31.1 & 2.5 & 34.4\\
FM Tau & 32.2 & 35.4 & 0 & 0 & 32.2 & 0 & 0.3 & 0 & 2.0 & 0.47 & 0\\ 
IP Tau & 42.9 & 25.8 & 0 & 6.9 & 22.3 & 0 & 2.1 & 0 & 3.0 & 2.8 & 8.8\\
GG Tau A & 42.3 & 14.1 & 11.3 & 12.7 & 16.9 & 0 & 2.3 & 0.6 & 1.7 & 3.4 & 28.8\\
GG Tau B & 44.4 & 0 & 8.9 & 0 & 24.4 & 13.3 & 8.9 & 0 & 2.5 & 29.4 & 11.8\\
Hen 3-600 A & 10.2 & 0 & 10.2 & 0 & 68.0 & 4.8 & 5.4 & 1.4 & 4.8 & 36.2 & 31.9\\
FN Tau & 30.2 & 0 & 21.8 & 0 & 28.5 & 16.8 & 2.0 & 0.8 & 4.0 & 27.4 & 30.4\\
V410 Anon 13 & 48.1 & 0 & 0 & 0 & 33.7 & 11.2 & 4.8 & 2.2 & 1.9 & 27.5 & 0\\
CY Tau & 37.3 & 0 & 6.2 & 6.2 & 31.1 & 11.2 & 5.0 & 3.1 & 2.3 & 27.9 & 18.0\\
\hline
\end{tabular}
\tablenotetext{a}{\footnotesize Optical constants for amorphous 
                 pyroxene Mg$_{0.8}$Fe$_{0.2}$SiO$_3$ 
                 from \citet{dor95}, 
                 assuming CDE2 \citep{fab01}}
\tablenotetext{b}{\footnotesize Optical constants for amorphous 
                 olivine MgFeSiO$_4$ from \citet{dor95}, 
                 assuming CDE2}
\tablenotetext{c}{\footnotesize Optical constants for amorphous 
                 pyroxene Mg$_{0.8}$Fe$_{0.2}$SiO$_3$ 
                 from \citet{dor95}, 
                 using the Bruggeman EMT and Mie theory
                 (Bohren \& Huffman 1983) 
                 with a volume fraction of vacuum of 
                 $f$\,=\,0.6 for porous spherical grains 
                 of radius 5$\mum$}
\tablenotetext{d}{\footnotesize Optical constants for 
                  amorphous olivine 
                  MgFeSiO$_4$ from \citet{dor95}, 
                  using the Bruggeman EMT and Mie theory
                 (Bohren \& Huffman 1983) 
                 with a volume fraction of vacuum of 
                 $f$\,=\,0.6 for porous spherical grains 
                 of radius 5$\mum$}            
\tablenotetext{e}{\footnotesize Optical constants for 
                  amorphous carbon 
                  (``ACAR'') from \citet{zub96}, assuming CDE2}
\tablenotetext{f}{\footnotesize Opacities for crystalline pyroxene 
                  Mg$_{0.9}$Fe$_{0.1}$SiO$_3$ from \citet{chi02}}
\tablenotetext{g}{\footnotesize Optical constants for 
                  3 crystallographic axes of forsterite, 
                  Mg$_{1.9}$Fe$_{0.1}$SiO$_4$, 
                  from \citet{fab01}, assuming CDE \citep{bh83}}
\tablenotetext{h}{\footnotesize Optical properties for $\alpha$ 
                  quartz from \citet{wc96}, assuming CDE}
\tablenotetext{i}{\footnotesize $\chi^{2}$/dof
                  (dof\,=\,8) is determined over 
                  8\,$<$\,$\lambda$\,$<$\,14$\mum$} 
\tablenotetext{j}{\footnotesize Percentage of crystalline dust
                  (pyroxene, forsterite and silica) compared to 
                   all silicates in the dust model.}
\tablenotetext{h}{\footnotesize Percentage of large silicate dust
                  (large amorphous pyroxene and large amorphous 
                  olivine) compared to all silicates in the dust model.}
}
\end{table}

\end{document}